\documentclass[%
twocolumn,
superscriptaddress,
 amsmath,amssymb,
 aps,prx
]{revtex4-2}

\usepackage{bbold}
\usepackage{amsfonts}
\usepackage{dsfont}
\usepackage{xcolor}
\usepackage{bm}
\usepackage{hyperref}
\hypersetup{plainpages=false,colorlinks=true,linkcolor=blue, citecolor=blue, urlcolor=blue}

\usepackage{xcolor}
\usepackage{bbold}
\usepackage{mathrsfs}
\usepackage{natbib}
\usepackage{here}

\usepackage{graphicx}

\newcommand{\calcium}{Ca$_5$Ir$_3$O$_{12}$ }

\usepackage{array,mathtools,amssymb,booktabs}
\newcolumntype{C}{>{$}c<{$}}

\begin{document}
\title{{\it Ab initio} Derivation of Low-Energy Hamiltonians for Systems with Strong Spin-Orbit Interaction and Its Application to \calcium}
\author{Maxime Charlebois} 
\affiliation{Waseda Research Institute for Science and Engineering, Waseda University, Okubo, Shinjuku, Tokyo, 169-8555, Japan.} 

\author{Jean-Baptiste Mor\'{e}e} 
\affiliation{Waseda Research Institute for Science and Engineering, Waseda University, Okubo, Shinjuku, Tokyo, 169-8555, Japan.} 

\author{Kazuma Nakamura}\email{kazuma@mns.kyutech.ac.jp}
\affiliation{Graduate School of Engineering, Kyushu Institute of Technology, Kitakyushu 804-8550, Japan}

\author{Yusuke Nomura}
\affiliation{RIKEN Center for Emergent Matter Science, 2-1 Hirosawa, Wako, Saitama 351-0198, Japan}

\author{Terumasa Tadano} 
\affiliation{Research Center for Magnetic and Spintronic Materials, National Institute for Materials Science, Tsukuba 305-0047, Japan}

\author{Yoshihide Yoshimoto}
\affiliation{Department of Computer Science, The University of Tokyo,7-3-1 Hongo, Bunkyo-ku, Tokyo 113-0033}

\author{Youhei Yamaji} 
\affiliation{Department of Applied Physics, The University of Tokyo, Hongo, Bunkyo-ku, Tokyo 113-8656, Japan}

\author{Takumi Hasegawa} 
\affiliation{Graduate School of Integrated Arts and Sciences, Hiroshima University, Higashihiroshima, Hiroshima 739-8521, Japan} 

\author{Kazuyuki Matsuhira}
\affiliation{Graduate School of Engineering, Kyushu Institute of Technology, Kitakyushu 804-8550, Japan}

\author{Masatoshi Imada} 
\affiliation{Waseda Research Institute for Science and Engineering, Waseda University, Okubo, Shinjuku, Tokyo, 169-8555, Japan.} 
\affiliation{Toyota Physical and Chemical Research Institute, Yokomichi, Nagakute, Aichi, 480-1118, Japan.} 

\begin{abstract}
We present an {\it ab initio} derivation method for effective low-energy Hamiltonians of material with strong spin-orbit interactions. The effective Hamiltonian is described in terms of the Wannier function in the spinor form, and effective interactions are derived with the constrained random phase approximation (cRPA) method. 
Based on this formalism and the developed code, we derive an effective Hamiltonian of a strong spin-orbit interaction material Ca$_5$Ir$_3$O$_{12}$. This system consists of three edge-shared IrO$_6$ octahedral chains arranged along the $c$ axis, and the three Ir atoms in the $ab$ plane compose a triangular lattice. For such a complicated structure, we need to set up the Wannier spinor function under the local coordinate system. We found that a density-functional band structure near the Fermi level is formed by local $d_{xy}$ and $d_{yz}$ orbitals. Then, we constructed the {\it ab initio}  $d_{xy}/d_{yz}$ model. The estimated nearest neighbor transfer $t$ is close to 0.2~eV, and the cRPA onsite $U$ and neighboring $V$ electronic interactions are found to be 2.4-2.5~eV and 1~eV, respectively. The resulting 
characteristic correlation strength defined by $(U-V)/t$ is above 7, and thus this material is classified as a strongly correlated electron system. The onsite transfer integral involved in the spin-orbit interaction is 0.2~eV, which is comparable to the onsite exchange integrals $\sim$ 0.2~eV, indicating that the spin-orbit-interaction physics would compete with the Hund physics. Based on these calculated results, we discuss possible rich ground-state low-energy electronic structures of spin, charge and orbitals with competing Hund, spin-orbit and strong correlation physics.
\end{abstract} 

\maketitle

\section{Introduction}
{\it Ab initio} studies of strongly correlated electron materials that allow quantitative comparisons with experimental results, and provide us with further predictive powers and materials design, are one of the grand challenges in materials science, where the conventional density functional theory (DFT) widely used for weakly correlated materials does not offer satisfactory accuracy. Instead, recent developments of the {\it ab initio} scheme with the reduction to effective low-energy Hamiltonians offer a promising framework~\cite{Aryasetiawan_2004,Imada_Miyake_2010}. For example, some of the high-$T_{\rm c}$ superconductors were studied in this  multi-scale {\it ab initio} scheme for correlated electrons (MACE), and successfully reproduced the experimental phase diagram quantitatively, one for a cuprate superconductor~\cite{Hirayama_2019,Ohgoe_2020} and the other for an iron-based superconductor~\cite{Miyake_2010,Misawa_2012,Misawa_2014}.

However, this scheme has not been examined extensively for the cases of coexisting electron correlations and spin-orbit interaction except for few cases as Refs.~\onlinecite{Arita_2012,Yamaji_2014}, while treating such interplay is coming increasingly important in various fields such as studies on topological materials. We focus on the efficient implementation of the spin-orbit interaction (SOI) to the above MACE scheme. Recently, an open source software called RESPACK~\cite{KazumaNakamura_2021,RESPACK_Nakamura_2016} was developed for the purpose of deriving effective low-energy Hamiltonians of strongly correlated material ranging from 
iron-based superconductors~\cite{cRPA_Nakamura_2008,Miyake_2010}, cuprates~\cite{cRPA_Hirayama_2018,cRPA_Hirayama_2019}, 
nickelates~\cite{cRPA_Nomura_2019,cRPA_Hirayama_2020}, 
3$d$ and 4$d$ transition-metal oxides~\cite{cRPA_Vaugier_2012}, 
magnetic insulators~\cite{cRPA_Huang_2020}, 
graphites~\cite{cRPA_Wehling_2011}, 
fullerides and aromatic compounds~\cite{cRPA_Nomura_2012,cRPA_Kim_2016}, 
organic compounds~\cite{cRPA_Nakamura_2009,cRPA_Nakamura_2012},  
zeolite systems~\cite{cRPA_Nakamura_2009_2,cRPA_Nohara_2011},
surface and interface systems~\cite{cRPA_Ersoy_2012,cRPA_Hansmann_2013,cRPA_Okamoto_2014,cRPA_Tadano_2019}, 
lanthanides~\cite{cRPA_Nilsson_2013,cRPA_Moree_2018}, and 
actinide dioxides~\cite{cRPA_Amadon_2014,cRPA_Seth_2017,cRPA_Moree_2021}. 
In addition, handling for band-entangled systems~\cite{cRPA_Miyake_2009,cRPA_Ersoy_2011} and development of an extension scheme to purely-low-dimensional systems
~\cite{cRPA_Nakamura_2010,cRPA_Nomura_2013} have also been made. Here, we also discuss an implementation of SOI into RESPACK to make it easier for the public use.  

As an application of the scheme, we derive effective low-energy Hamiltonians of Ca$_5$Ir$_3$O$_{12}$, whose enigmatic experimental properties wait for theoretical support. Electrons on Ir 5$d$ orbitals in Ca$_5$Ir$_3$O$_{12}$ are expected to have strong SOI, and we elucidate its effects on the level of the effective Hamiltonians. 

Ca$_5$Ir$_3$O$_{12}$ has a hexagonal structure (${\it P\bar{6}2m}$)~\cite{Wakeshima_2003} with quasi-one-dimensional Ir-atom chain forming the triangular lattice structure in the direction perpendicular to the chain. It shows insulating (or semiconducting) transport properties presumably in variable range hopping regime~\cite{Wakeshima_2003,Cao_2007}, while the Ir $5d$ orbitals 
are partially filled in the DFT electronic structure even in the presence of SOI~\cite{Matsuhira_2018} as we clarify later, which suggests a substantial role of the electron correlation. In fact, the compound shows the Curie-Weiss susceptibility~\cite{Cao_2007} at high temperatures and undergoes two phase transitions, one at around 105 K and the other at 7.8 K. The latter was identified as the magnetic transition from the susceptibility, specific heat~\cite{Wakeshima_2003,Cao_2007} and $\mu$SR~\cite{Franke_2011} measurements, while the structure of the magnetic order is not identified yet. The magnetism below 7.8 K is possibly characterized by canted antiferromagnetism or spin glass because of the different response of the magnetic susceptibility to the field and zero-field cool measurements applied parallel to the $c$-axis. On the other hand, the mechanism of the 108 K transition is also actively studied in terms of the phonon property analysis by using the Raman~\cite{Hasegawa_2020} and inelastic X-ray scattering~\cite{Hanate_2020} measurements, but the structural deformation is so small that it has not been identified.

The DFT electronic structure suggests the quasi-1D-like electronic dispersion~\cite{Matsuhira_2018}, which also implies that disorder plays a key role in transport properties. Iridium valence is furthermore fluctuating nominally between $1/3$ of Ir$^{4+}$ ($d^{9}$, namely effective spin $1/2$ hole) and $2/3$ of Ir$^{5+}$ ($d^8$, namely spin-1 hole) counted from the closed shell of $t_{2g}$ orbitals ($d^{6}$) in the octahedron environment formed by the oxygen, suggesting the role of charge fluctuations and/or disproportionation. Since the unit cell contains three Ir atoms with triangular lattice composed of three quasi-one-dimensional chains, a naive expectation would be the disproportionation into periodic alignment of two Ir$^{5+}$ and one Ir$^{4+}$ atoms, which is expected to dissolve the magnetic frustration arising from the triangular structure of Ir atoms. However, so far, as mentioned above, the crystal symmetry was claimed to stay the same through the two transitions from the Raman~\cite{Hasegawa_2020} and X-ray~\cite{Hanate_2020} spectra, and the unit cell stays the same where the three Ir atoms are all apparently equivalent~\cite{Wakeshima_2003} at least above 15 K. It also shows the nonlinear conductivity under strong electric fields~\cite{Matsuhira_2018}. All of these involved but attractive features require theoretical support to clarify the origin of the bad metal behavior. 

In this paper, we derive an effective low-energy Hamiltonian of Ca$_5$Ir$_3$O$_{12}$ using RESPACK considering the SOI.
From the analysis of the DFT band structure, we found that $d_{xy}$ and $d_{yz}$ orbitals are essential in describing the low-energy electronic structure near the Fermi level, and we then derived the $d_{xy}/d_{yz}$ Hamiltonian. An electron transfer structure with strong one-dimensional anisotropy was confirmed, but on the other hand, it was found that the details of the interchain electron transfer can affect the low-energy electronic structure. In addition, the 
correlation strength, $(U-V)/t$ of this material is very large, about 7, and it was found that the electronic correlation is relevant to the low-energy physics. Furthermore, since the magnitude of the SOI is nearly the same as the size of the exchange integral, the SOI physics and the Hund physics compete with each other. We discuss possible spin-charge orders that can be expected from the analysis of the derived parameters for the $d_{xy}/d_{yz}$ Hamiltonian.

The present paper is organized as follows: In Section~\ref{sec_method}, we give definition of an effective Hamiltonian to be derived and describe calculation details of the Wannier function in the spinor form and how to obtain the reducible $k$-point wave function data from the irreducible $k$-point data. Implementation details within~
RESPACK and calculation conditions are also given in this section. In Section~\ref{sec_result}, we show the DFT~
band structure, the calculated 
Wannier functions, and the derived 
effective-model parameters for 
the~
spin-orbit material Ca$_5$Ir$_3$O$_{12}$. Discussion and summary for possible ground-state electronic structures of this material are given in Section~\ref{sec_summary_discussion}. In appendix~\ref{app:Hderivation-KN}, we give the details of the effective Hamiltonian to be derived and, in Appendix~\ref{app:J-and-K}, we show the characteristic aspect of the exchange integrals in the spinor formalism. Also, we discuss in Appendix~\ref{app:SingleChainHamiltonian} mathematical aspects of single chain Hamiltonian based on our derived parameters.

\section{Method}\label{sec_method} 

\subsection{Effective Hamiltonian to be derived}\label{effective_Hamiltonian} 

In the present study, we consider a derivation of an effective low-energy Hamiltonian consisting of $2N_w$ Wannier orbitals, where $N_w$ is the total number of Kramers pairs~
in the~
unit cell. We suppose that the first $N_w$ Wannier orbitals have mainly an up-spin component, and the last~
$N_w$ Wannier orbitals have mainly a down-spin component. Then, we introduce indices $i$ and $\sigma$ to specify the Wannier spinor state, where $i$ is the index for the Wannier orbital and $\sigma$ specifies one of the two components in the Kramers pair. The effective Hamiltonian to be considered is written within the two-center integrals as 
\begin{eqnarray}
\mathcal{H}
&=& 
\sum_{{\bf R} {\bf R}'} \sum_{ij} \sum_{\sigma\rho} 
t_{i{\bf R} j{\bf R}'}^{\sigma\rho}
a_{i\sigma{\bf R}}^{\dagger} a_{j\rho{\bf R}'}  \nonumber \\
&+& \frac{1}{2} \sum_{{\bf R} {\bf R}'} \sum_{ij}  \sum_{\sigma\rho} 
 U_{i {\bf R} j {\bf R}'}^{\sigma\rho} 
 a_{i\sigma{\bf R}}^{\dagger} a_{j\rho{\bf R}'}^{\dagger} 
 a_{j\rho{\bf R}'} a_{i\sigma{\bf R}} \nonumber \\ 
&+& \frac{1}{2} \sum_{{\bf R} {\bf R}'} \sum_{ij} \sum_{\sigma\rho} 
 J_{i {\bf R} j {\bf R}'}^{\sigma\rho}  
 a_{i\sigma{\bf R}}^{\dagger} a_{j\rho{\bf R}'}^{\dagger} 
 a_{i\rho{\bf R}} a_{j\sigma{\bf R}'} \nonumber \\ 
&+& \frac{1}{2} \sum_{{\bf R} {\bf R}'} \sum_{ij} \sum_{\sigma\rho} 
 K_{i {\bf R} j {\bf R}'}^{\sigma\rho}  
 a_{i\sigma{\bf R}}^{\dagger} a_{i\rho{\bf R}}^{\dagger} 
 a_{j\rho{\bf R}'} a_{j\sigma{\bf R}'}.
\label{eq:H}                
\end{eqnarray} 
This Hamiltonian is based on the ``colinear approximation" on the $\sigma$ and $\rho$ degrees of freedom, and a detailed derivation is presented in Appendix~\ref{app:Hderivation-KN}. In Eq. (\ref{eq:H}), $a_{i\sigma{\bf R}}^{\dagger}$ and $a_{i\sigma{\bf R}}$ are creation and annihilation operators, respectively, of an electron in the $(i\sigma)$-th Wannier state at a lattice ${\bf R}$. The creation operator is defined by $|\Phi_{i\sigma{\bf R}}\rangle=a_{i\sigma{\bf R}}^{\dagger}|0\rangle$ and $\Phi_{i\sigma{\bf R}}({\bf r})$ is the Wannier function in the spinor form as 
\begin{eqnarray}
  \Phi_{i\sigma{\bf R}}({\bf r})   
 = \left(\begin{array}{cc} 
    \phi_{i\sigma{\bf R}}^{u}({\bf r}) \vspace{0.3cm} \\ 
    \phi_{i\sigma{\bf R}}^{d}({\bf r})  \\ 
  \end{array} \right) 
\label{spinor-wannier} 
\end{eqnarray}
and
\begin{eqnarray}
  \Phi_{i\sigma{\bf R}}^{\dagger}({\bf r})
= \biggl(\phi_{i\sigma{\bf R}}^{u*}(\mathbf{r}) \ \phi_{i\sigma{\bf R}}^{d*}(\mathbf{r}) \biggr). 
\label{spinor-wannier-d} 
\end{eqnarray}
In the spinor representation, it is necessary to pay attention to the quantization axis for describing the components. In this formulation, the Wannier spinor function is represented along the Cartesian-$z$ axis. $\phi({\bf r})$ is the spatial component along this axis, and the superscript ``$u$" and ``$d$" describe the up and down components, respectively. It should be noted here that the subscript $\sigma$ specifies the spin components along the local quantization axis (or front or back degree of freedom of the Kramers~
pair). 

$t_{i{\bf R} j{\bf R}'}^{\sigma\rho}$ in Eq.~(\ref{eq:H}) is a transfer integral as 
\begin{eqnarray}
t_{i{\bf R} j{\bf R}'}^{\sigma\rho}
=\int_V d{\bf r}\ 
\Phi_{i\sigma{\bf R}}^{\dagger}({\bf r}) 
\mathcal{H}_{{\rm KS}}({\bf r}) 
\Phi_{j\rho{\bf R}'}({\bf r}) 
\label{tij} 
\end{eqnarray}
with
\begin{eqnarray}
 \mathcal{H}_{{\rm KS}} = \left(\begin{array}{cc} 
  {\cal H}^{uu} & {\cal H}^{ud}  \vspace{0.3cm} \\ 
  {\cal H}^{du} & {\cal H}^{dd} \\ \end{array} \right),  
\end{eqnarray}
and it is evaluated as the Wannier matrix element of the Kohn-Sham (KS) Hamiltonian. The integral of the right-hand side in Eq.~(\ref{tij}) is taken over the crystal volume $V$. Also, $U_{i {\bf R} j {\bf R}'}^{\sigma\rho}$, $J_{i {\bf R} j {\bf R}'}^{\sigma\rho}$, and $K_{i {\bf R} j {\bf R}'}^{\sigma\rho}$ in Eq.~(\ref{eq:H}) are static interaction integrals defined as 
\begin{widetext}
\begin{eqnarray}
U_{i {\bf R} j {\bf R}'}^{\sigma\rho} 
=\lim_{\omega\to0}\int_V d{\bf r} \int_V d{\bf r}'
\Phi_{i\sigma{\bf R}}^{\dagger}({\bf r}) 
\Phi_{i\sigma{\bf R}}({\bf r}) 
W({\bf r,r}',\omega) 
\Phi_{j\rho{\bf R}'}^{\dagger}({\bf r}') 
\Phi_{j\rho{\bf R}'}({\bf r}'),
\label{Wij}
\end{eqnarray}
\begin{eqnarray}
J_{i {\bf R} j {\bf R}'}^{\sigma\rho} 
= \lim_{\omega\to0} \int_V d{\bf r} \int_V d{\bf r}'
\Phi_{i\sigma{\bf R}}^{\dagger}({\bf r}) 
\Phi_{j\sigma{\bf R}'}({\bf r}) 
W({\bf r,r}',\omega) 
\Phi_{j\rho{\bf R}'}^{\dagger}({\bf r}') 
\Phi_{i\rho{\bf R}}({\bf r}'),
\label{Jij}
\end{eqnarray}
and 
\begin{eqnarray}
K_{i {\bf R} j {\bf R}'}^{\sigma\rho}  
= \lim_{\omega\to0} \int_V d{\bf r} \int_V d{\bf r}'
\Phi_{i\sigma{\bf R}}^{\dagger}({\bf r}) 
\Phi_{j\sigma{\bf R}'}({\bf r}) 
W({\bf r,r}',\omega) 
\Phi_{i\rho{\bf R}}^{\dagger}({\bf r}') 
\Phi_{j\rho{\bf R}'}({\bf r}'), 
\label{Kij}
\end{eqnarray}
\end{widetext} 
respectively. $W({\bf r,r}',\omega)$ is a frequency-dependent effective Coulomb interaction calculated within the constrained random phase approximation (cRPA)~\cite{Aryasetiawan_2004} 
 or constrained GW approximation (cGW)~\cite{Hirayama2013,Hirayama2017} and this is a scalar form. In the spinor formalism, the Wannier function is not real, so, in principle, $J_{i {\bf R} j {\bf R}'}^{\sigma\rho} \ne K_{i {\bf R} j {\bf R}'}^{\sigma\rho}$. We discuss the structure of the $J$ and $K$ matrices in detail in Appendix~\ref{app:J-and-K}. In the application to Ca$_5$Ir$_3$O$_{12}$ in this paper, we employ the cRPA.

\subsection{Initial guess for maximally localized Wannier functions in the spinor formalism} \label{initial_guess_for_wannier} 
The Wannier function is obtained from a transformation of the Bloch function as 
\begin{eqnarray}
    \Phi_{i\sigma{\bf R}}(\mathbf{r})
   =\frac{1}{\sqrt{N_k}}\sum_{{\bf k}}^{N_k} \sum_{n} A_{n,i\sigma}^{{\bf k}} 
    \Psi_{n{\bf k}}({\bf r})
    e^{-i{\bf k}\cdot{\bf R}}, 
\end{eqnarray} 
where ${\bf k}$ is a wavevector in the $k$-mesh, $N_k$ is the total number of the $k$ mesh, and \{$A_{n,i\sigma}^{{\bf k}}$\} is a transformation matrix from the Bloch basis to the Wannier basis.~
As well as the Wannier function $\Phi_{i\sigma{\bf R}}(\mathbf{r})$, the Bloch function $\Psi_{n{\bf k}}({\bf r})$ is also represented in the spinor format as 
\begin{eqnarray} 
    \Psi_{n{\bf k}}({\bf r})=
    \left( \begin{array}{c} 
    \psi_{n{\bf k}}^{u}({\bf r}) \vspace{0.3cm} \\   
    \psi_{n{\bf k}}^{d}({\bf r}) \\ 
    \end{array} \right),  
\end{eqnarray}
where $\psi({\bf r})$ is the spatial component along the Cartesian-$z$ axis. To specify the matrix \{$A_{n,i\sigma}^{{\bf k}}$\}, we utilize the maximally-localized Wannier-function algorithm~\cite{Marzari_1997,Souza_2001}. 

In the spinor formalism, we construct~
$2N_w$ Wannier functions. A proper setting of the initial guess for the Wannier function is important for obtaining stable results. In particular, in the present complex oxide Ca$_5$Ir$_3$O$_{12}$, it is required to prepare an initial guess in a form in which the coordinate system describing the Wannier function is set to a local coordinate symmetry based on a local octahedron IrO$_6$ (see Sec.~\ref{sec_structure}). Also, the proper SU(2) rotation of the spinor quantization axis is important for keeping the symmetry of the matrix elements of the transfer and interaction $U$, $J$, and $K$ matrices.

In the construction of the practical Wannier function, we calculate a projection of the initial guess onto the Bloch function; 
we first prepare the following initial guesses for the Wannier functions at the home cell ${\bf R}={\bf 0}$; for the spinors with $\sigma = \uparrow$, 

\begin{eqnarray}
  \Phi_{i\uparrow{\bf 0}}^{ini}({\bf r})   
 = \left(\begin{array}{cc} 
    \phi_{i\uparrow{\bf 0}}^{ini,u}({\bf r}) \vspace{0.3cm} \\ 
    \phi_{i\uparrow{\bf 0}}^{ini,d}({\bf r}) \\ 
  \end{array} \right)
 = \mathbf{U}(\bm{\alpha}) 
 \left(\begin{array}{cc} 
    g^{L}_{i}({\bf r})  \vspace{0.3cm} \\ 
    0 \\ 
  \end{array} \right),
  \label{wannier-first-half}
\end{eqnarray}
and, for the spinors with $\sigma = \downarrow$, 
\begin{eqnarray}
  \Phi_{i\downarrow{\bf 0}}^{ini}({\bf r})   
 = \left(\begin{array}{cc} 
    \phi_{i\downarrow{\bf 0}}^{ini,u}({\bf r}) \vspace{0.3cm} \\ 
    \phi_{i\downarrow{\bf 0}}^{ini,d}({\bf r}) \\ 
  \end{array} \right)
 = \mathbf{U}(\bm{\alpha}) 
 \left(\begin{array}{cc} 
    0  \vspace{0.3cm} \\ 
    g^{L}_{i}({\bf r}) \\ 
  \end{array} \right). 
   \label{wannier-latter-half}
\end{eqnarray}
$g^{L}_{i}({\bf r})$ is the Gaussian function including $s$-, $p$-, or $d$-type. It should be noted here that the Gaussian $g^{L}_{i}({\bf r})$ orientation is aligned to the local coordinate system~\cite{KazumaNakamura_2021}; 
even if $g^{L}_{i}({\bf r})$ is expressed in the Cartesian coordinate system $\bf r$. So there is a spatial rotation between the usual orbital definition and $g^{L}_{i}({\bf r})$. In the present material, Ca$_5$Ir$_3$O$_{12}$, $g^{L}_i({\bf r})$ in Eqs.~(\ref{wannier-first-half}) and (\ref{wannier-latter-half}) is a $d$-type Gaussian function aligned to the local coordinates centered to the IrO$_6$ octahedron, and is taken to be common for the Kramers pair of $\Phi_{i\uparrow{\bf 0}}^{ini}({\bf r})$ and $\Phi_{i\downarrow{\bf 0}}^{ini}({\bf r})$. 

On top of that, the spinor in the right hand side in Eqs.~(\ref{wannier-first-half}) and (\ref{wannier-latter-half}) is expressed in the quantization axis along the $z$ direction of the local coordinate system; 
at this stage, the minor component is set to zero and the major component is set to $g^{L}_i({\bf r})$. 
The Bloch function obtained from the band structure calculation is conventionally expressed in the quantization axis along the Cartesian-$z$ direction. Then, to evaluate the inner product of the Bloch function and the initial guess, we need a rotation of the quantization axis of the initial guess to the Cartesian-$z$ direction.  
The spinor components along the Cartesian-$z$ quantization axis are obtained by applying the ${\bf U}({\bm \alpha}$) to the spinor along the local-$z$ quantization axis, as it is done in Eqs.~(\ref{wannier-first-half}) and (\ref{wannier-latter-half}). Here ${\bf U}({\bm \alpha}$) is the SU(2) matrix characterized by the Euler angles $\bm{\alpha} = (\alpha,\beta,\gamma)$, which rotates the quantization axis from the local frame to the Cartesian frame. We should note that this ${\bf U}({\bm \alpha}$) is related to the SO(3) matrix $\mathbf{L(\bm \alpha)}$ describing the spatial rotation from the Cartesian coordinate system to the local one. The SO(3) matrix $\mathbf{L}(\bm \alpha)$ is  
\begin{eqnarray}
 \mathbf{L}(\bm \alpha) = \bigr( \mathbf{L}_x\ \mathbf{L}_y\ \mathbf{L}_z \bigl) 
 =
 \left(\begin{array}{ccc} 
  L_{Xx} & L_{Xy} & L_{Xz} \\ 
  L_{Yx} & L_{Yy} & L_{Yz} \\ 
  L_{Zx} & L_{Zy} & L_{Zz} \\ 
  \end{array} \right), 
  \label{mat_local} 
\end{eqnarray}
where ${\bf L}_x$, ${\bf L}_y$, and ${\bf L}_z$ are unit vectors along the local $x$, $y$, and $z$ directions, respectively. Note that for the matrix $\mathbf{L}(\bm \alpha)$ to be SO(3), hence a pure rotation, we must define the unit vector $\mathbf{L}_x$, $\mathbf{L}_y$ and $\mathbf{L}_z$ to be orthogonal. The orientation of $\mathbf{L}_z$ is the most important. Also, $X$, $Y$, and $Z$ are the Cartesian coordinates. The matrix elements in Eq.~(\ref{mat_local}) can be calculated from atomic positions forming the local octahedron IrO$_6$. We note that  Ca$_5$Ir$_3$O$_{12}$ includes three IrO$_6$ octahedron units and each IrO$_6$ octahedron has its own different local direction (see Sec.~\ref{sec_structure}). The SU(2) $\bf{U}(\bm{\alpha})$ is
\begin{eqnarray}
 \bf{U}({\bm \alpha}) = 
  \left(\begin{array}{cc} 
    \exp\bigl(-i\frac{\alpha+\gamma}{2}\bigr)\cos\bigl(\frac{\beta}{2}\bigr) &
   -\exp\bigl(-i\frac{\alpha-\gamma}{2}\bigr)\sin\bigl(\frac{\beta}{2}\bigr) \vspace{0.3cm} \\ 
    \exp\bigl(i\frac{\alpha-\gamma}{2}\bigr)\sin\bigl(\frac{\beta}{2}\bigr)  & 
    \exp\bigl(i\frac{\alpha+\gamma}{2}\bigr)\cos\bigl(\frac{\beta}{2}\bigr) \\
  \end{array} \right). \nonumber \\ 
  \label{mat_Ualpha}
\end{eqnarray}
The Euler angles $\bm{\alpha} = (\alpha,\beta,\gamma)$ in Eq.~(\ref{mat_Ualpha}) are obtained as follows: The local coordinates ${\bf L}({\bm \alpha})$ in Eq.~(\ref{mat_local}) are 
expressed by using the Euler angles as:
\begin{widetext}
\begin{eqnarray}
  \mathbf{L}(\bm \alpha) =
 \left(\begin{array}{ccc} 
  \cos\alpha\cos\beta\cos\gamma-\sin\alpha\sin\gamma & -\cos\alpha\cos\beta\sin\gamma -\sin\alpha\cos\gamma  & \cos\alpha\sin\beta \\ 
  \sin\alpha\cos\beta\cos\gamma+\cos\alpha\sin\gamma & -\sin\alpha\cos\beta\sin\gamma + \cos\alpha\cos\gamma & \sin\alpha\sin\beta \\ 
 -\sin\beta\cos\gamma & \sin\beta\sin\gamma & \cos\beta \\ 
  \end{array} \right).
  \label{Euler} 
\end{eqnarray}
\end{widetext}

Through the comparison of the above equation with Eq.~(\ref{mat_local}), we obtain the Euler angles as follows: 
\begin{eqnarray}
 \alpha&=&\arctan\biggl(\frac{L_{Yz}}{L_{Xz}}\biggr)\ {\rm or}\ \arctan\biggl(\frac{L_{Yz}}{L_{Xz}}\biggr)+\pi, \label{alpha} \\
 \beta&=&\arccos({L_{Zz}})\ {\rm or}\ \arccos({L_{Zz}})+\pi, \label{beta} \\
 \gamma&=&\arctan\biggl(-\frac{L_{Zy}}{L_{Zx}}\biggr)\ {\rm or}\ \arctan\biggl(-\frac{L_{Zy}}{L_{Zx}}\biggr)+\pi. \label{gamma} 
\end{eqnarray} 
In the numerical calculation by {\sc Fortran}, since the $\arctan$ and $\arccos$ functions return the values of [$\pi/2, \pi/2$] and [$0, \pi$], respectively, as mod functions, there is an ambiguity of a multiple of $\pi$  from the divisor. Thus, we look for a suitable Euler angle, taking into account this ambiguity. On top of that, when $\sin\beta$ in Eq.~(\ref{Euler}) is zero, the notorious Gimbal lock problem appears. In this case, we determine the Euler angles as follows:
\begin{eqnarray}
 \alpha&=&\arctan\biggl(\frac{L_{Yx}}{L_{Xx}}\biggr)\ {\rm or}\ \arctan\biggl(\frac{L_{Yx}}{L_{Xx}}\biggr)+\pi, \label{alpha-GL} \\
 \beta&=&0\ {\rm or}\ \pi, \label{beta-GL} \\
 \gamma&=&0. \label{gamma-GL} 
\end{eqnarray}
By inserting Eqs.~(\ref{alpha}), (\ref{beta}), and (\ref{gamma}) or (\ref{alpha-GL}), (\ref{beta-GL}), and (\ref{gamma-GL}) into Eq.~(\ref{mat_Ualpha}), we obtain the SU(2) matrix ${\bf U}({\bm \alpha})$.  

After employing the initial guesses in Eqs.~(\ref{wannier-first-half}) and (\ref{wannier-latter-half}), we calculate the~
initial matrix $A_{n,i\sigma}^{{\bf k},ini}$ for a given Bloch function $\Psi_{n{\bf k}}$ as the following inner product
\begin{eqnarray}
     A_{n,i\sigma}^{{\bf k},ini} 
 &=& \langle \Psi_{n{\bf k}} | \Phi^{ini}_{i\sigma{\bf 0}} \rangle \nonumber \\ 
 &=& \int_V \bigl( \psi_{n{\bf k}}^{u*}(\mathbf{r}) 
  \ \psi_{n{\bf k}}^{d*}(\mathbf{r}) \bigr)
     \left(\begin{array}{cc} 
       \phi_{i\sigma{\bf 0}}^{ini,u}({\bf r}) \vspace{0.3cm}  \\ 
       \phi_{i\sigma{\bf 0}}^{ini,d}({\bf r}) \\ 
     \end{array} \right) d\mathbf{r} \nonumber \\ 
 &=& \int_V \psi_{n{\bf k}}^{u*}(\mathbf{r}) \phi_{i\sigma{\bf 0}}^{ini,u}({\bf r}) d\mathbf{r}
  +  \int_V \psi_{n{\bf k}}^{d*}(\mathbf{r}) \phi_{i\sigma{\bf 0}}^{ini,d}({\bf r}) d\mathbf{r}. \nonumber \\ 
\end{eqnarray} 
The rest of the calculation procedure is basically the same as the conventional procedure; we perform the spillage and spread minimization based on the maximally-localized Wannier-function algorithm~\cite{Marzari_1997,Souza_2001}. 

\subsection{Relation between the irreducible and reducible Bloch functions}
In the practical calculation, there is another important technical point. It is a procedure to generate reducible $k$-point wavefunctions from irreducible $k$-point wavefunctions. In usual band calculation, only the irreducible $k$-point wavefunctions are calculated and stored. Therefore, in the Wannier function~
and RPA/cRPA calculations that follow the band calculation, the wavefunctions of the reducible $k$ points must be generated from the wavefunctions of the irreducible $k$ points. This technique is important for reducing computation cost and memory size. 

Let us consider the relationship between irreducible and reducible wavefunctions. The reducible wave function is expanded by plane waves as 
\begin{eqnarray}
    \Psi_{n{\bf k}}(\mathbf{r})
    \!&=&\!\left( \begin{array}{c} 
    \psi_{n{\bf k}}^{u}({\bf r}) \vspace{0.3cm} \\   
    \psi_{n{\bf k}}^{d}({\bf r}) \\ 
    \end{array} \right) \nonumber \\ 
    \!&=&\! \sum_{{\bf G}} \left( \begin{array}{c} 
    C^{\ u}_{{\bf G}n}({\bf k}) \vspace{0.3cm} \\   
    C_{{\bf G}n}^{\ d}({\bf k}) \end{array} \right) 
    \frac{\!\exp \bigl[ +i({\bf k}\!+\!{\bf G})\!\cdot\!\mathbf{r} \bigr]}
    {\sqrt{\Omega}},   
      \label{spinor-wf} 
\end{eqnarray}
where ${\bf k}$ is reducible $k$ point, ${\bf G}$ is reciprocal lattice vector for expansion of the wave function at the reducible $k$ point, and $\Omega$ is the unit-cell volume.
$C_{{\bf G}n}^{u}({\bf k})$ and $C_{{\bf G}n}^{d}({\bf k})$ are the expansion coefficients, which are expressed along the Cartesian-$z$ axis. These are written with coefficients at irreducible $k$ point, ${\bf k}^*$, as follows:  
\begin{eqnarray}
 \left( \begin{array}{c} 
  C^{\ u}_{{\bf G}n}({\bf k}) \vspace{0.3cm} \\  
  C^{\ d}_{{\bf G}n}({\bf k}) \\ \end{array} \right) 
 \!=\!\bm{S}({\bm \alpha}^{\prime}) \left( \begin{array}{c} 
  C_{{\bf G}^*n}^{\ u}({\bf k}^*)
  e^{-i({\bf G}\!+\!{\bm \Delta}_{rw})\cdot{\mathcal T}}  \vspace{0.3cm} \\
  C_{{\bf G}^*n}^{\ d}({\bf k}^*)
  e^{-i({\bf G}\!+\!{\bm \Delta}_{rw})\cdot{\mathcal T}} \\ \end{array} \right) \nonumber \\ 
  \label{spinor-coef} 
\end{eqnarray}
with 
\begin{eqnarray}
{\bf G}^*={\mathcal R}^{-1}({\bf G}+{\bm \Delta}_{rw}).
\label{Gstar}
\end{eqnarray}
Here, ${\bf G}^*$ is the reciprocal lattice vector for expansion of the wave function at the irreducible $k$ point. ${\mathcal R}$ is a rotation matrix representing the rotational operation for the system, and ${\mathcal T}$ is partial translation vector. It should be noted here that the ${\mathcal R}$ matrix operates in a reciprocal space, and converts a vector ${\bf k}^*+{\bf G}^*$ into a vector ${\bf k}+{\bf G}$ as  
\begin{eqnarray}
{\bf k}+{\bf G} = {\mathcal R} ({\bf k}^*+{\bf G}^*).   
\end{eqnarray}
${\bm \Delta}_{rw}$ in Eqs.~(\ref{spinor-coef}) and (\ref{Gstar}) is a rewind vector which is introduced to pull back the $k$ point after the rotational operation, ${\mathcal R}{\bf k}^*$, to the inside of the Brillouin zone. In the practical calculation, we first look for ${\bf G}^*$ that satisfies Eq.~(\ref{Gstar}), and then specify $C_{{\bf G}^*n}^{\ u}({\bf k}^*)$ and $C_{{\bf G}^*n}^{\ u}({\bf k}^*)$ in the right-hand side of Eq.~(\ref{spinor-coef}). $\bm{S}(\bm{\alpha}^{\prime})$ in Eq.~(\ref{spinor-coef}) is an SU(2) matrix rotating the spinor, which can be evaluated as~
described in Sec.~\ref{initial_guess_for_wannier}; the ${\mathcal R}$ is conventionally represented in the basic reciprocal lattice coordinates. Therefore, we convert ${\mathcal R}$ into ${\mathcal R}^{XYZ}$ which is a rotation matrix in the Cartesian coordinates as 
\begin{eqnarray}
{\mathcal R}^{XYZ}={\bf B}{\mathcal R}{\bf B}^{-1} 
\end{eqnarray}
with 
\begin{eqnarray}
 {\bf B} =  \bigr( \mathbf{b}_1\ \mathbf{b}_2\ \mathbf{b}_3 \bigl) 
 =
 \left(\begin{array}{ccc} 
  b_{X1} & b_{X2} & b_{X3} \\ 
  b_{Y1} & b_{Y2} & b_{Y3} \\ 
  b_{Z1} & b_{Z2} & b_{Z3} \\ 
  \end{array} \right). 
  \label{mat_B} 
\end{eqnarray}
Thus, with ${\mathcal R}^{XYZ}$, we determine the Euler angles ($\alpha^{\prime}$, $\beta^{\prime}$, $\gamma^{\prime}$) along the same procedure as in Eq.~(\ref{alpha}), (\ref{beta}), and (\ref{gamma}) or (\ref{alpha-GL}), (\ref{beta-GL}), and (\ref{gamma-GL}), for ${\bf S}({\bm \alpha}^{\prime})$ of Eq.~(\ref{spinor-coef}). Since the spinor is a polar vector, we have to extract the pure rotational part in the symmetry operation. So, we first evaluate the determinant $\det\|{\mathcal R}^{XYZ}\|$, and, if $\det\|{\mathcal R}^{XYZ}\|<0$, we define $\tilde{{\mathcal R}}^{XYZ}=-{\mathcal R}^{XYZ}$ and determine the Euler angles from this $\tilde{{\mathcal R}}^{XYZ}$ matrix. Finally, by using the resulting Euler angles, we evaluate the SU(2) rotation matrix ${\bf S}({\bm \alpha}^{\prime})$ of Eq.~(\ref{spinor-coef}) to obtain $C_{{\bf G}n}^{\ u}({\bf k})$ and $C_{{\bf G}n}^{\ d}({\bf k})$. 

\subsection{Implementation in RESPACK and calculation conditions} 
\label{sec_implcalc}
The method mentioned above was implemented in RESPACK~\cite{KazumaNakamura_2021} which is a first-principles calculation software for evaluating the interaction parameters of materials and is able to calculate maximally localized Wannier functions, response functions based on the RPA and related optical properties, and frequency-dependent electronic interaction parameters. RESPACK supports band-calculation codes using norm-conserving pseudopotentials with plane-wave basis sets, and automatic generation scripts for converting the band-calculation results to the RESPACK inputs are prepared for xTAPP~\cite{Yamauchi_1996} and {\sc Quantum ESPRESSO}~\cite{QE-2009,QE-2017} packages.  

Density-functional band structure~
calculations for Ca$_5$Ir$_3$O$_{12}$ are performed by using xTAPP and {\sc Quantum ESPRESSO} with the experimental crystal structure with lattice parameters~\cite{Wakeshima_2003}: $a$ = 9.3491 ${\rm \AA}$ and $c$ = 3.1713 ${\rm \AA}$. We use the Perdew--Burke--Ernzerhof  type~\cite{Perdew_1996} for the exchange-correlation functional. In {\sc Quantum ESPRESSO}, the norm-conserving pseudopotentials are generated by the code ONCVPSP (Optimized Norm-Conserving Vanderbilt PSeudopotential)~\cite{Hamann_2013}, and~
are obtained from the PseudoDojo~\cite{Setten_2018}. In xTAPP, the norm-conserving pseudopotentials~\cite{Kleinman_1982,Troullier_1991} are generated as follows: The Ir pseudopotential is constructed for both of the valence and semicore electronic configurations. For the former pseudopotential, we consider a slightly ionic configuration of $(5d)^7(6s)^1(6p)^0$, where the core-electron configuration is (Xe)$(4f)^{14}$; the $4f$ electrons are frozen and excluded from the pseudopotential. The cutoff radius for the local potential $r_{loc}$ is 1.7 bohr, and those for the non-local $s$, $p$, and $d$ projectors are 2.1, 2.4, and 2.1 bohr, respectively. We apply the partial-core correction with a cutoff radius $r_{pcc}$ of 1.3 bohr. Also, the semicore-type pseudopotential of Ir was construncted for an ionic semicore configuration of $(5s)^{2}(5p)^{6}(5d)^{7}(6s)^1(6p)^0$ with $r_{loc}$ = 1.0 bohr. The cutoff radii for the non-local $s$, $p$, and $d$ projectors are 1.0, 1.0, and 1.2 bohr
, respectively, where the $5s$ and $6s$ channels and the $6s$ and $6p$ channels share the same cutoff radius. The Ca pseudopotential is constructed for a slightly ionic $(3s)^2(3p)^6(4s)^{1.6}(4p)^{0.3}(3d)^0$ configuration with $r_{loc}$ = 1.0 bohr. The cutoff radius of the non-local $s$, $p$, and $d$ projectors are 1.0 bohr. The O pseudopotential is generated with $r_{loc}$ = 1.0 bohr for a valence configuration of $(2s)^2(2p)^4$. The cutoff radii of the non-local $s$, $p$, and $d$ projectors are 1.0, 1.0, and 1.0 bohr, respectively. 

We use 8$\times$8$\times$8 $k$-points for sampling in the first Brillouin zone. The energy cutoff is set to be 144 Ry for the wave functions and 576 Ry for the charge density. The Fermi energy in the band calculations was estimated with the broadening techniques with the smearing of 0.0272~eV~\cite{Methfessel_1989} (for the calculations with {\sc Quantum Espresso}, the Gaussian smearing of the same value was used). The interaction parameters are calculated using the cRPA method~\cite{Aryasetiawan_2004,cRPA_Nakamura_2008,cRPA_Ersoy_2011}, in which we employ the band disentanglement scheme~\cite{cRPA_Ersoy_2011}. The energy cutoff for the dielectric function is set to be 20 Ry. The total number of bands used in the polarization calculation is 340 for the calculations with the valence-type Ir pseudopotential and 404 for the semicore-type one, which includes the unoccupied states up to $\sim$ 29~eV with respect to the Fermi level. The integral over the Brillouin zone is calculated with the generalized tetrahedron technique~\cite{Fujiwara_2003,Nohara_2009} with a smearing of 0.1~eV. 

To study the SOI effect on the electronic structure, we perform the usual GGA calculation and compare the results with the SOI. We call the former calculation as GGA and the latter calculation as SO-GGA. In SO-GGA, the Wannier functions was constructed in a band-select mode that constructs the Wannier function by directly specifying the Bloch bands related to the Wannier function without setting the energy window. On the other hand, the GGA Wannier function were constructed by specifying the inner and outer energy window. The inner window was set to [$-$0.32~eV, 0.75~eV] for both xTAPP and {\sc Quantum Espresso}, where the energy zero is the Fermi level. The outer window was taken to be [$-$0.61~eV, 0.75~eV] for both xTAPP and {\sc Quantum Espresso}. The unoccupied states up to $\sim$ 26~eV with respect to the Fermi level are included in the polarization calculation. 

\section{Results}\label{sec_result}
\subsection{Crystal structure}\label{sec_structure}
\begin{figure*}[htpb] 
\begin{center} 
\includegraphics[width=0.8\textwidth]{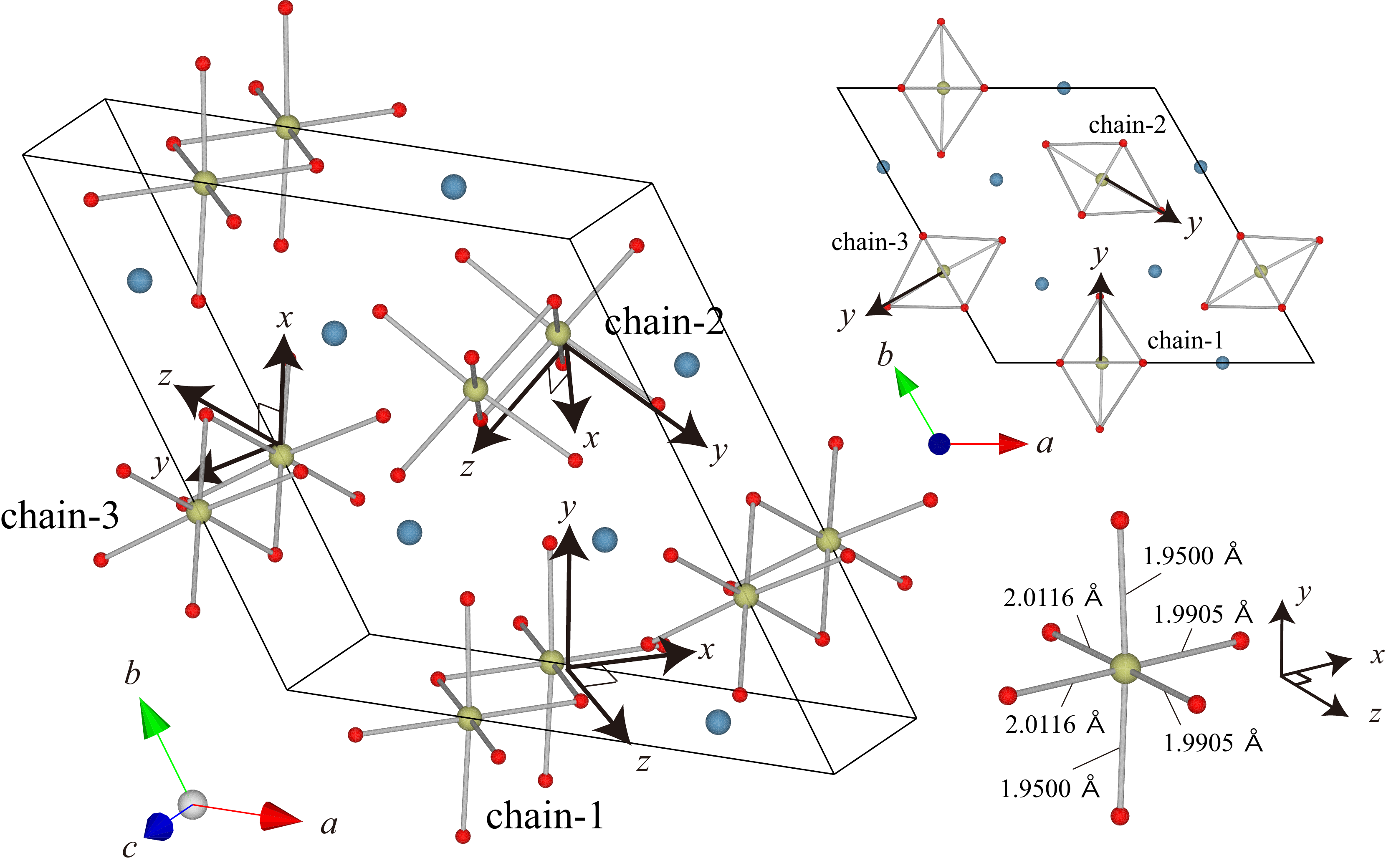}
\end{center} 
\vspace{-0.5cm} 
\caption{Crystal structure of Ca$_5$Ir$_3$O$_{12}$, where Ca, O, and Ir atoms are depicted by blue, small-red, and yellow spheres, respectively (drawn by VESTA~\cite{Momma_2011}). The unit cell contains three edge-shared IrO$_6$ chains along the $c$ axis, and these chains are denoted as chain-1, chain-2, and chain-3. The local coordinate system based on the IrO$_6$ octahedron in each chain are also depicted. We note that the local $y$ axes are in the $ab$ plane, and the local coordinates have a rotational symmetry of 120 degree rotation around the $c$ axis. To show this symmetry clearly, we give in the top right inset the cross-section from the $c$-axis direction. 
We also show in the bottom right inset the IrO$_6$ octahedron including the Ir-O bond length data, from which we see that the structure is considerably distorted and has no inversion symmetry.} 
\label{geometry_Ca5Ir3O12}
\end{figure*}

Figure~\ref{geometry_Ca5Ir3O12} shows the crystal structure of Ca$_5$Ir$_3$O$_{12}$. The dark-blue, light-yellow, and small-red spheres indicate Ca, Ir, and O atoms, respectively, and bonds are drawn between the Ir and O atoms. This material consists of three budding rods of edge-shared IrO$_6$ octahedra in the unit cell. These IrO$_6$ rods are aligned along the $c$-axis, and these are related by a $120^{\circ}$ rotation symmetry around the $c$ axis. We refer to each rod as chain-$n$ with $n$ being an index for the chains and running from 1 to 3. Following the octahedral convention, a local coordinate system associated with the octahedron is defined as shown in Fig.~\ref{geometry_Ca5Ir3O12}. The local $y$ axes are taken in the $ab$ plane and in the direction of the vertex oxygen of the IrO$_6$ octahedron. The local coordinates of each chain also match by a $120^{\circ}$ rotation about the $c$ axis (see the top right inset of Fig.~\ref{geometry_Ca5Ir3O12}). We also note that the IrO$_6$ octahedron is distorted and has no inversion symmetry (see the bottom right inset of Fig.~\ref{geometry_Ca5Ir3O12}).    

\subsection{Band calculation} 
\begin{figure}[htpb] 
\begin{center} 
\includegraphics[width=0.475\textwidth]{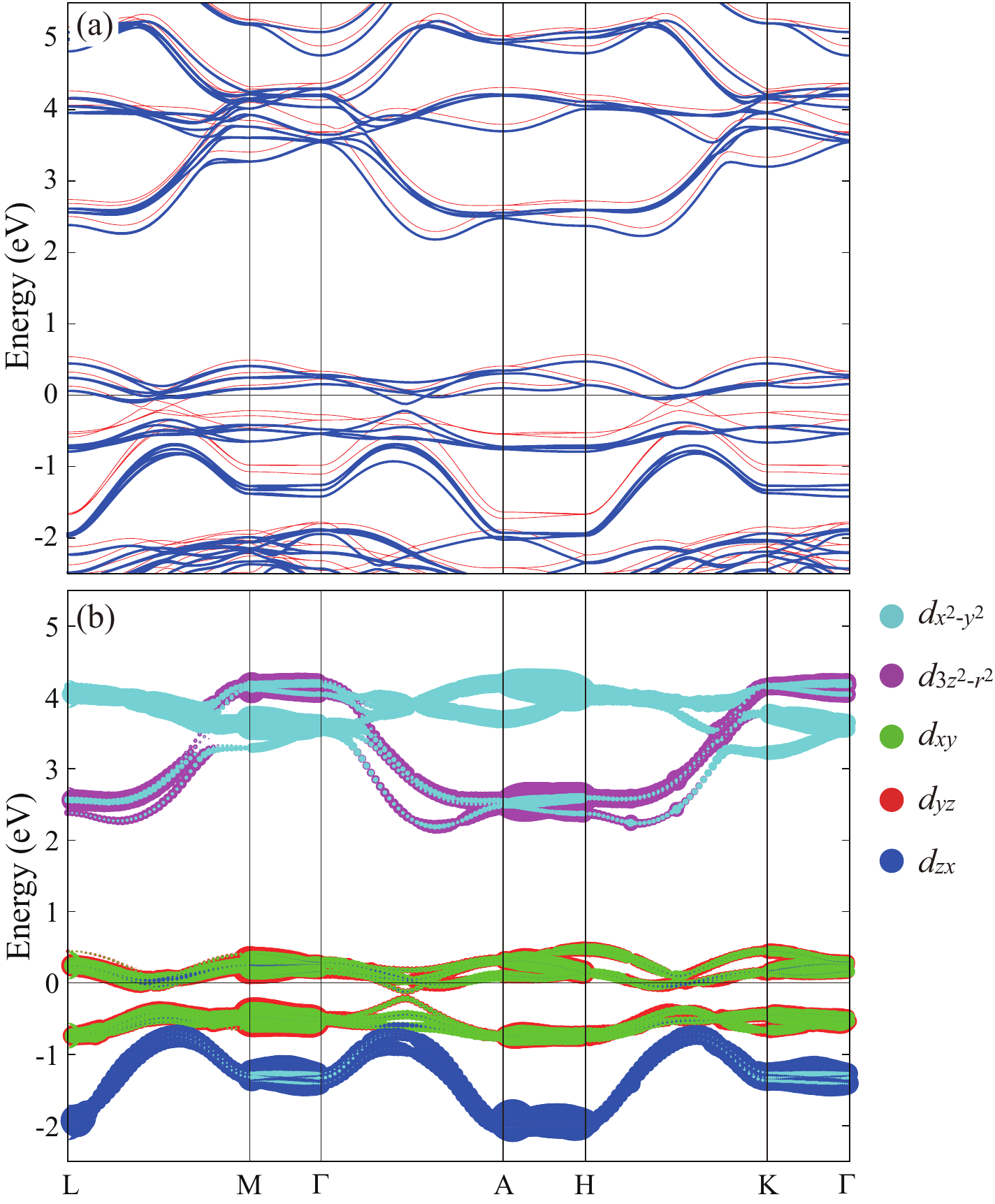}
\end{center} 
\vspace{-0.5cm} 
\caption{(a) {\it Ab initio} density functional band structure of Ca$_5$Ir$_3$O$_{12}$. Thick-blue and thin-red curves are the results with and without the spin-orbit interaction, respectively. The energy zero is the Fermi level. Dispersions are plotted along the high-symmetry points, where $\Gamma$=(0, 0, 0), L=(0, $b^{*}$/2, $c^{*}$/2), M=(0, $b^{*}$/2, 0), A=(0, 0, $c^{*}$/2), H=($-a^{*}$/3, 2$b^{*}$/3, $c^{*}$/2), and K=($-a^{*}$/3, 2$b^{*}$/3, 0) with $a^{*}, b^{*}$, and $c^{*}$ being basic vectors of the reciprocal lattice, respectively. (b) Fat band results for 5$d$ orbital of Ir atom.} 
\label{band_5d}
\end{figure}

We show in Fig.~\ref{band_5d} (a) calculated GGA and SO-GGA band structures of Ca$_5$Ir$_3$O$_{12}$ by thin-red and thick-blue curves, respectively. From the comparison, we see that the SOI affects the electronic structure near the Fermi level, especially leading to the band split and gap opening in the GGA bands along the L-M and H-K lines.

To see the character of the global band structure, we show in Fig.~\ref{band_5d}(b) results of the fat-band analysis for the SO-GGA band, where the band structure is decomposed into the 5$d$ orbitals of the Ir atoms. The bands near 2~eV to 4~eV are composed by $d_{x^2-y^2}$ and $d_{3z^2-r^2}$ orbitals, while the bands around $-$2~eV to 0.5~eV consist of $d_{xy}$, $d_{yz}$, and $d_{zx}$ orbitals. This band structure results from the crystal field splitting of 5$d$ orbitals into $e_g$ and $t_{2g}$ groups under the crystal field of oxygen atoms at the corner of an octahedron. Since the IrO$_6$ octahedron is distorted, the degeneracy within each group is lifted. From these results, we found that the low-energy bands around the Fermi level consist of mainly $d_{xy}$ and $d_{yz}$ orbitals. It should be noted here that the $d$ orbitals follow the local coordinate. The reason why the $xy$ and $yz$ bands have the higher energy than the $zx$ band is that the distance from the Ir atom to the apex oxygen in the local $y$ direction (1.95 \AA) is shorter than that to the in-plane oxygen (1.99 and 2.01 \AA). Thus, in the present study, we consider a derivation of an effective Hamiltonian for the $d_{xy}$ and $d_{yz}$ orbitals, and call this the $d_{xy}/d_{yz}$ Hamiltonian. Since there are three Ir atoms in the unit cell, this model is composed of the 12 Wannier spinor states.  

\subsection{Onsite energy diagram} 
\begin{figure}[htpb] 
\begin{center} 
\includegraphics[width=0.425\textwidth]{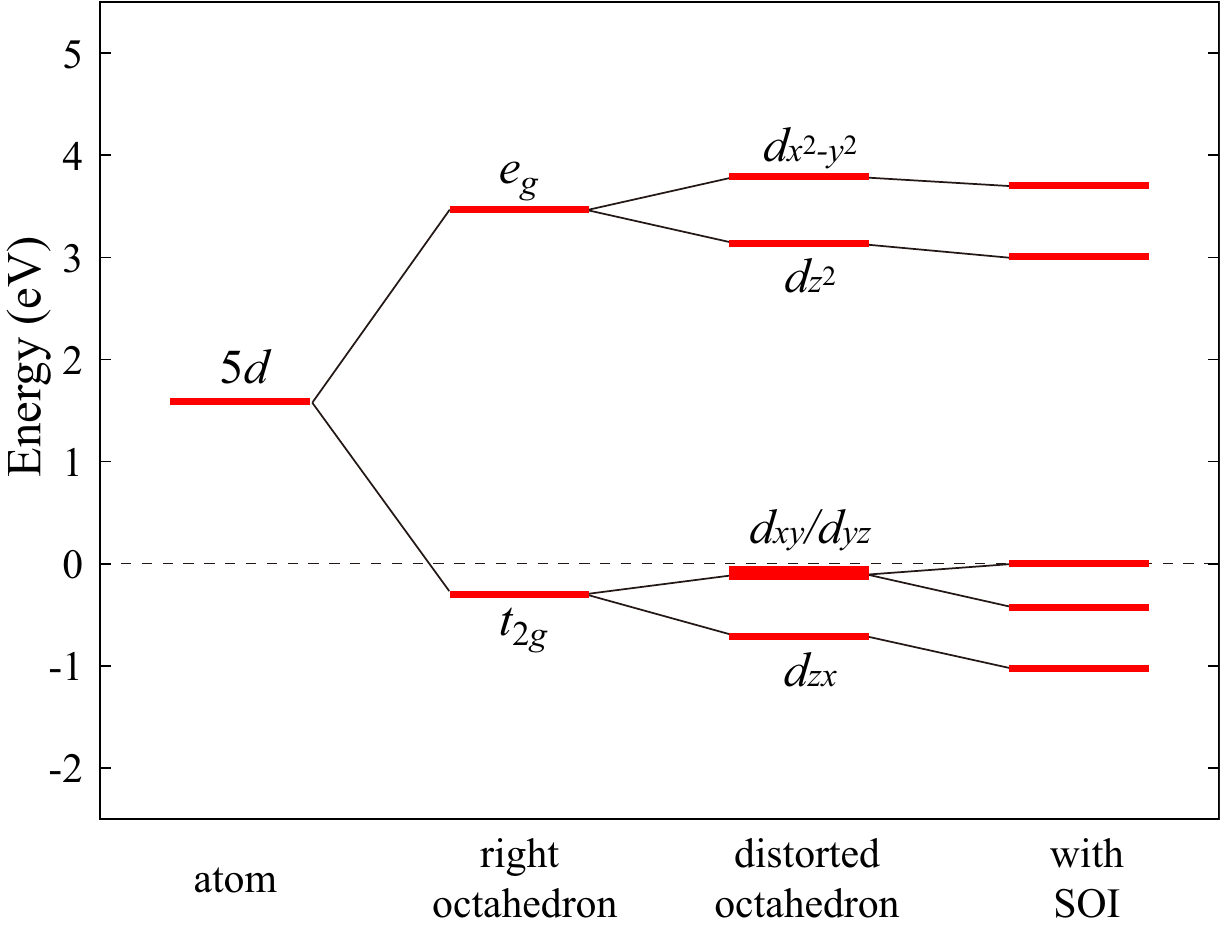}
\end{center} 
\vspace{-0.5cm} 
\caption{Level diagram for onsite energy of Ir. By considering the crystal field splitting as large as 3.8~eV, the 5$d$ level of Ir splits the $e_g$ and $t_{2g}$ levels. Further, due to the distortion of the IrO$_6$ octahedron mentioned in the bottom right inset of Fig.~\ref{geometry_Ca5Ir3O12}, the $e_g$ and $t_{2g}$ levels undergo a small band splitting near 0.6~eV. We note that this distortion hardly causes a level splitting for $d_{xy}/d_{yz}$ orbitals; the level deviation between the $d_{xy}$ and $d_{yz}$ orbitals is about 0.071~eV. When the spin-orbit interaction acts on these levels, the $d_{xy}$ and $d_{yz}$ levels near the Fermi level split.} 
\label{onsite-energy}
\end{figure}

Based on Fig.~\ref{band_5d}, the on-site energy diagram of Ir is summarized in Fig.~\ref{onsite-energy}. The crystal field splitting of the 5$d$ level of iridium is 3.764~eV, and further, due to the distortion of the octahedron, both $e_g$ and $t_{2g}$ levels undergo a small band splitting near 0.6~eV. We note that this distortion cannot cause an appreciable level splitting between $d_{xy}$ and $d_{yz}$ orbitals; the splitting of the two orbits is as small as 0.071~eV. Finally, when the SOI acts on these levels, the $d_{xy}$ and $d_{yz}$ levels further split (see below).  

\subsection{$d_{xy}/d_{yz}$ Hamiltonian} 
\begin{figure}[htpb] 
\begin{center} 
\includegraphics[width=0.475\textwidth]{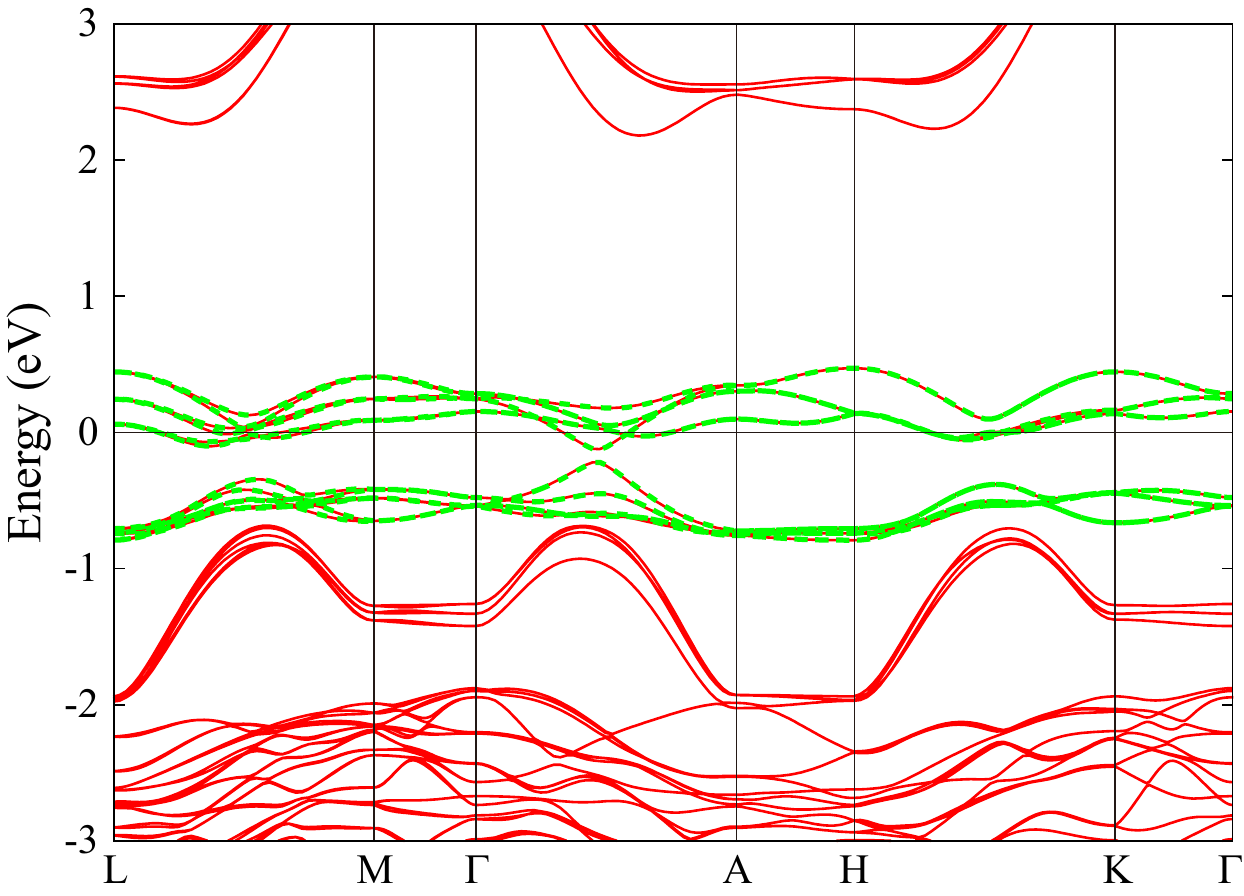}
\end{center} 
\vspace{-0.5cm} 
\caption{Comparison of the Wannier-interpolation band (green-dashed curves)  based on the $d_{xy}$ and $d_{yz}$ orbitals and the original SO-GGA band (red-solid curves). The view of the figure is the same as that of Fig.~\ref{band_5d} (a).} 
\label{band_dxydyz}
\end{figure}

From here, we derive the $d_{xy}/d_{yz}$ Hamiltonian, because these two bands are well isolated from other bands near the Fermi level as one sees in Fig.~\ref{band_5d}. Figure~\ref{band_dxydyz} is a comparison of the Wannier-interpolation band (green-dashed curves) and the original SO-GGA band (red-solid curves). We see a good agreement between the two bands. We note that the initial guess setting is important for the present \calcium; we set the $d_{xy}$ and $d_{yz}$ Gaussian orbitals as initial guesses, where the $d$ orbitals are represented in the local coordinates within each IrO$_6$ octahedra (see Fig.~\ref{geometry_Ca5Ir3O12}). Also, the initial guesses are represented as a pure spin up or down state along the local quantization axis [Eqs.~(\ref{wannier-first-half}) and (\ref{wannier-latter-half})]. With this treatment, all twelve Wannier functions have the same spread. This setting is very important to keep the right three-fold symmetry of matrix elements in the Hamiltonian [transfer \{$t_{i {\bf R} j {\bf R}'}^{\sigma\rho}$\} and interaction \{$U_{i {\bf R} j {\bf R}'}^{\sigma\rho}$\}, \{$J_{i {\bf R} j {\bf R}'}^{\sigma\rho}$\}, and \{$K_{i {\bf R} j {\bf R}'}^{\sigma\rho}$\} matrices in Eq.~(\ref{eq:H})] by the calculated Wannier functions. 

\subsection{Maximally localized Wannier function} 
\begin{figure*}[htpb] 
\begin{center} 
\includegraphics[width=0.78\textwidth]{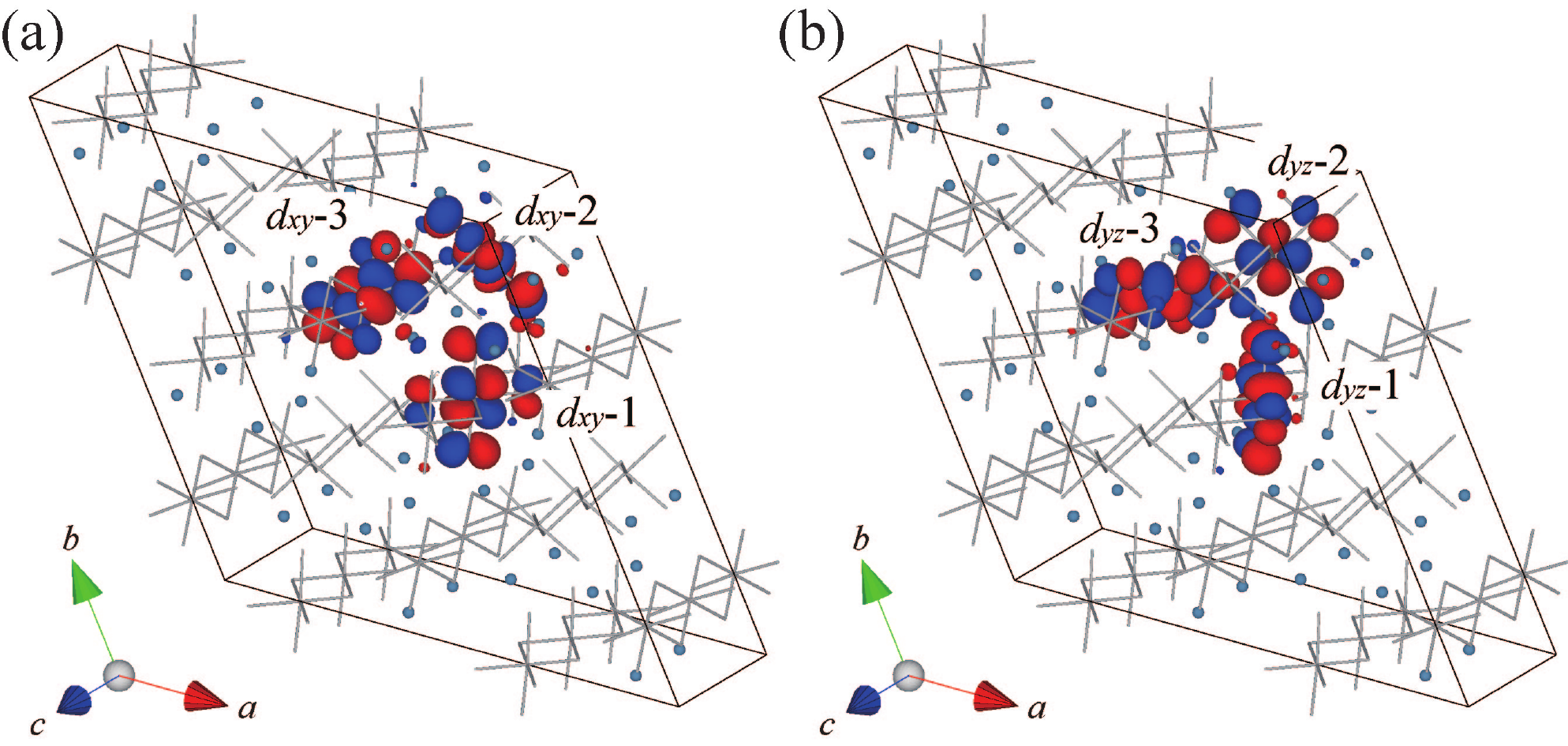}
\end{center} 
\vspace{-0.5cm} 
\caption{Visualization of maximally localized Wannier functions of Ca$_5$Ir$_3$O$_{12}$: The $d_{xy}$ (a) and $d_{yz}$ (b) Wannier functions are plotted as Re[$\bar{\phi}_{xy\uparrow{\bf 0}}^{u}({\bf r})$] and Re[$\bar{\phi}_{yz\uparrow{\bf 0}}^{u}({\bf r})$], respectively, in Eq.~(\ref{wannier-local}) (drawn by VESTA~\cite{Momma_2011}). The Wannier functions are drawn within the spatial range of a 2$\times$2$\times$2 supercell. In the figures, the octahedral IrO$_6$ chains are drawn by silver sticks, and Ca atoms are displayed by small-blue spheres. The $d_{xy}$ and $d_{yz}$ Wannier functions of the $n$th IrO$_6$ chain are described as $d_{xy}$-$n$ and $d_{yz}$-$n$, respectively. 
We note that the drawn three Wannier functions in the panels are independent and displayed on one plot.} 
\label{visualized_wannier}
\end{figure*}

We next describe details of the maximally localized Wannier functions of the $d_{xy}/d_{yz}$ Hamiltonian. The calculated real-space Wannier functions are displayed in Fig.~\ref{visualized_wannier}. The panels (a) and (b) illustrate the $d_{xy}$ and $d_{yz}$ Wannier functions, respectively. In this plot, three independent Wannier functions are shown in one panel together. This plot was made as follows: 
\begin{enumerate}
    \item First, we convert the resulting Wannier spinor represented along the Cartesian-$z$ axis into that in the local-$z$ axis as
    \begin{eqnarray}
      \left(\begin{array}{cc} 
       \bar{\phi}_{i\sigma{\bf 0}}^{u}({\bf r}) \vspace{0.3cm} \\ 
       \bar{\phi}_{i\sigma{\bf 0}}^{d}({\bf r})  \\ 
       \end{array} \right)
     = \mathbf{U}(\bm{\alpha})^{-1} 
       \left(\begin{array}{cc} 
       \phi_{i\sigma{\bf 0}}^{u}({\bf r})  \vspace{0.3cm} \\ 
       \phi_{i\sigma{\bf 0}}^{d}({\bf r}) \\ 
       \end{array} \right), 
    \label{wannier-local}
    \end{eqnarray}
    where $\mathbf{U}(\bm{\alpha})$ is the SU(2) matrix introduced in Eqs.~(\ref{wannier-first-half}) and (\ref{wannier-latter-half}). Also, $\bar{\phi}^{u}({\bf r})$ and $\bar{\phi}^{d}({\bf r})$ are the spatial components along the local-$z$ axis. With this conversion, we found that the components of the Wannier spinor concentrate on the major part. In the present compound, about 98 \% of the components are concentrated on the major part.
    \item Then, we plot the real part of the major part in Fig.~\ref{visualized_wannier}. Namely we plot Re[$\bar{\phi}_{xy\uparrow{\bf 0}}^{u}({\bf r})$] in the panel (a) and Re[$\bar{\phi}_{yz\uparrow{\bf 0}}^{u}({\bf r})$] in the panel (b). We see that the resulting Wannier functions are equivalent for the $d_{xy}$ and $d_{yz}$ orbitals.  
\end{enumerate}

\subsection{Transfer parameters}
\begin{figure}[htpb] 
\begin{center} 
\includegraphics[width=0.55\textwidth]{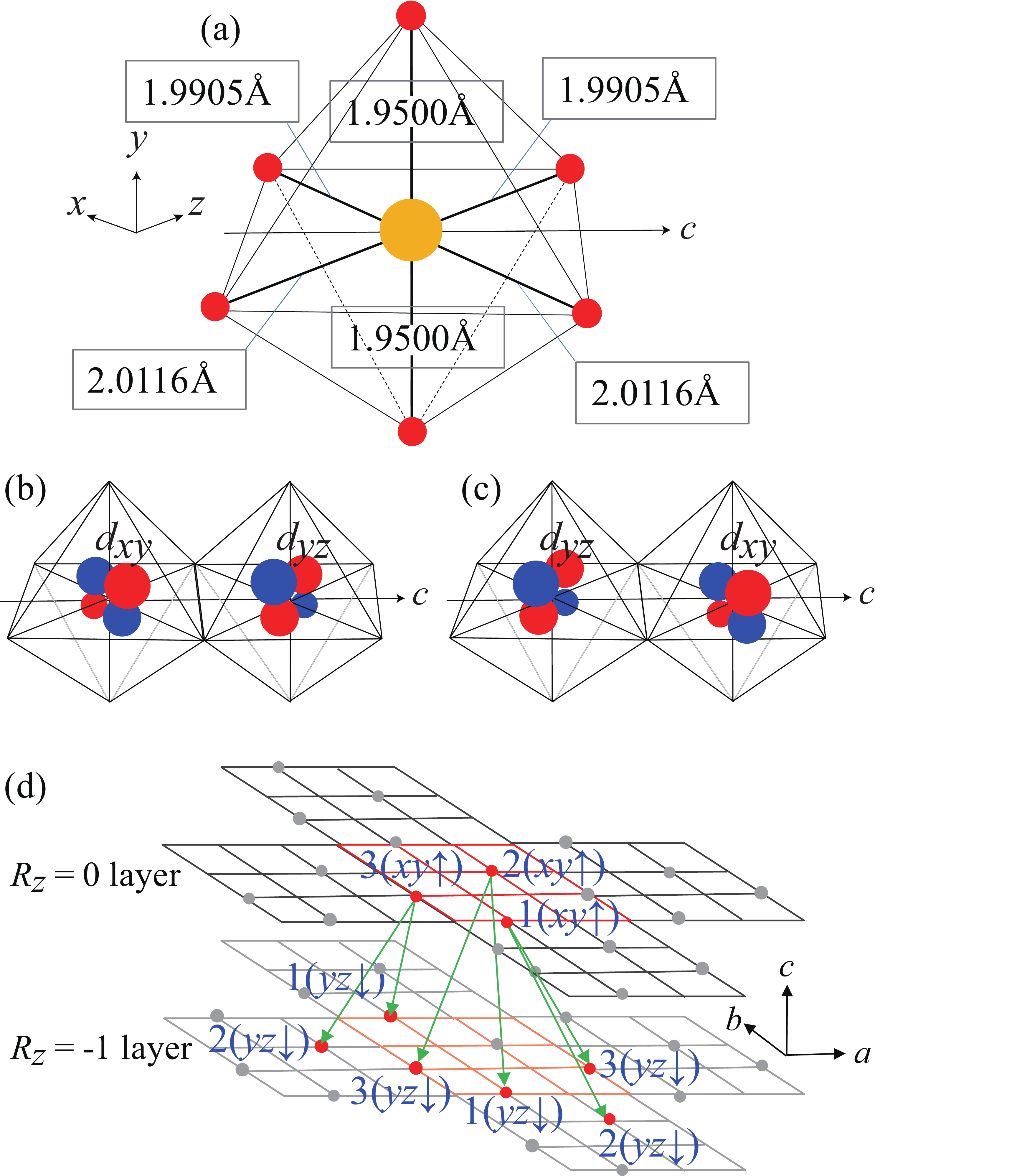}
\end{center} 
\vspace{-0.5cm} 
\caption{Schematic figure to understand configurations for main transfers in TABLE~\ref{transfer}: (a) Distorted IrO$_6$ octahedron including the Ir-O bond length data and the local coordinates are drawn in the left side. Panels (b) and (c) respectively describe nearest-neighbor ($NN$) pairs of the $d_{xy}$-$d_{yz}$ and $d_{yz}$-$d_{xy}$ transfers on the edge-sheared octahedrons along the $c$-axis. Due to the octahedral distortion, the transfer-path length via the bridging O site is appreciably longer in (b) than (c), leading to the difference in the transfer parameters ($t_{xy\uparrow,yz\uparrow}^{NN} \sim$ 0.18~eV and $t_{yz\uparrow,xy\uparrow}^{NN} \sim$ 0.22~eV). The panel (d) shows configurations for the largest interchain electron transfers (green arrows). The numbers in the panel denote Ir sites and types of orbitals are specified in the parentheses. We note that the largest interchain electron transfer occurs between Ir sites located at the nearest neighbor $ab$ plane each other and not along the $c$-axis but one Ir atom apart from that along the $c$-axis; it is larger than interchain electron transfers in the same $ab$ plane.} 
\label{trasfer_definition}
\end{figure}

Next, we discuss the one-body part of the $d_{xy}/d_{yz}$ Hamiltonian. TABLE~\ref{transfer} summarizes the main transfers of this model. The nearest neighbor ($NN$) transfers $t_{xy\uparrow,yz\uparrow}^{NN}$ and $t_{yz\uparrow,xy\uparrow}^{NN}$ are 0.178~eV and 0.215~eV, respectively. A schematic figure of the two neighboring orbitals (to understand these transfers defined by the bond between these two orbitals) is given in Fig.~\ref{trasfer_definition}, where the panel (a) displays a distorted octahedron, and panels (b) and (c) are two edge-shared octahedrons along the $c$-axis. The former displays the configuration of $t_{xy\uparrow,yz\uparrow}^{NN}$, and the latter is that of $t_{yz\uparrow,xy\uparrow}^{NN}$. The difference in the two transfers ($t_{xy\uparrow,yz\uparrow}^{NN}$ = 0.178~eV and $t_{yz\uparrow,xy\uparrow}^{NN}$ = 0.215~eV) can be understood in terms of the path-length difference in the transfer configurations due to the octahedron distortion. The $NN$ transfers between the same orbitals $t_{xy\uparrow,xy\uparrow}^{NN}$ and $t_{yz\uparrow,yz\uparrow}^{NN}$ are as small as 0.03~eV. The onsite transfer $t_{xy\uparrow,yz\uparrow}^{\rm onsite}$ is also as small as 0.027~eV. The largest interchain ($IC$) electron transfer $t_{IC}^{\rm largest}$ is 0.032~eV, whose schematic figure is depicted in Fig.~\ref{trasfer_definition} (d). It should be noted here that this transfer occurs between the pair located each in the neighboring $ab$ planes, where the pair partner is not located at the Ir atom along the $c$ axis (interplane nearest pair), but located at the the nearest neighbor Ir atom of the interplane nearest Ir atom; it is larger than the $IC$ transfer in the same $ab$ plane (nearly 0.028~eV). Lastly, the spin-orbit coupling $t_{xy\uparrow,yz\downarrow}^{\rm onsite}$ is remarkably large as 0.213~eV. 

The most interesting and important point is that the transfer parameters of SO-GGA and GGA are almost the same, and the only difference is due to the SOI matrix element $t_{xy\uparrow,yz\downarrow}^{\rm onsite}$. Looking at the comparison between the SO-GGA and GGA bands in Fig.~\ref{band_5d} (a), we notice that the two-band structures are different near the Fermi level, and the origin of this difference is obviously the SOI. TABLE~\ref{transfer}  also shows the difference between xTAPP and {\sc Quantum Espresso} results. Although a small difference is found in the values of the $NN$ transfers, we confirmed that the band dispersion of xTAPP is in perfect agreement with that of {\sc Quantum Espresso} at low energies.  
\begin{table}[htpb] 
 \caption{Main transfer parameters of the $d_{xy}/d_{yz}$ Hamiltonian of Ca$_5$Ir$_3$O$_{12}$, which are estimated as the matrix elements of the Kohn-Sham Hamiltonian in Eq.~(\ref{tij}) with respect to the maximally localized Wannier functions. In this table, we compare the SO-GGA and GGA results, and the second and third columns are results based on the xTAPP band calculation, and the fourth and fifth columns contain results with the {\sc Quantum Espresso} band calculation. We show 4 nearest-neighbor ($NN$) transfers along the chain ($c$-axis), onsite transfers, and absolute value of the largest interchain ($IC$) electron transfer. Definition for $t_{xy\uparrow,yz\uparrow}^{NN}$ and $t_{yz\uparrow,xy\uparrow}^{NN}$ is given in Figs.~\ref{trasfer_definition} (b) and (c), respectively. Also, the configuration for the largest $IC$ electron transfer $t_{IC}^{\rm largest}$ is drawn in Fig.~\ref{trasfer_definition} (d). $t_{xy\uparrow,yz\uparrow}^{\rm onsite}$ is nonzero because of the local coordinate defined along the distorted octahedron. The bottom $t_{xy\uparrow,yz\downarrow}^{\rm onsite}$ is the matrix element due to the onsite spin-orbit interaction. The unit of transfer integral is eV.}  
\begin{center} 
\begin{tabular}{l@{\ \ \ }r@{\ \ \ }r@{\ \ \ }r@{\ \ \ }r@{\ \ \ }r} 
 \hline \hline \\ [-5pt]
  & \multicolumn{2}{c}{xTAPP} & & \multicolumn{2}{c}{{\sc Quantum Espresso}} \\ 
    \cmidrule{2-3}                 \cmidrule{5-6} 
                                      & SO-GGA  & GGA   & & SO-GGA   & GGA   \\ \hline \\ [-5pt] 
 $t_{xy\uparrow,yz\uparrow}^{NN}$     & 0.178   & 0.173 & & 0.182    & 0.174 \\ [+5pt] 
 $t_{yz\uparrow,xy\uparrow}^{NN}$     & 0.215   & 0.209 & & 0.219    & 0.210 \\ [+5pt] 
 $t_{xy\uparrow,xy\uparrow}^{NN}$     & 0.027   & 0.024 & & 0.025    & 0.024 \\ [+5pt] 
 $t_{yz\uparrow,yz\uparrow}^{NN}$     & 0.027   & 0.024 & & 0.025    & 0.024 \\ [+5pt] 
 $t_{xy\uparrow,yz\uparrow}^{\rm onsite}$ & $-$0.027 & $-$0.036 & & $-$0.033 & 0.045 \\ [+5pt] 
 $|t_{IC}^{{\rm largest}}|$           & 0.032   & 0.041 & & 0.032    & 0.041 \\ [+5pt]
 $t_{xy\uparrow,yz\downarrow}^{\rm onsite}$ & 0.213 & - & & 0.215    & -     \\ [+5pt] 
 %
 %
 \hline \hline
\end{tabular} 
\end{center} 
\label{transfer} 
\end{table}

\begin{figure}[htpb] 
\begin{center} 
\includegraphics[width=0.475\textwidth]{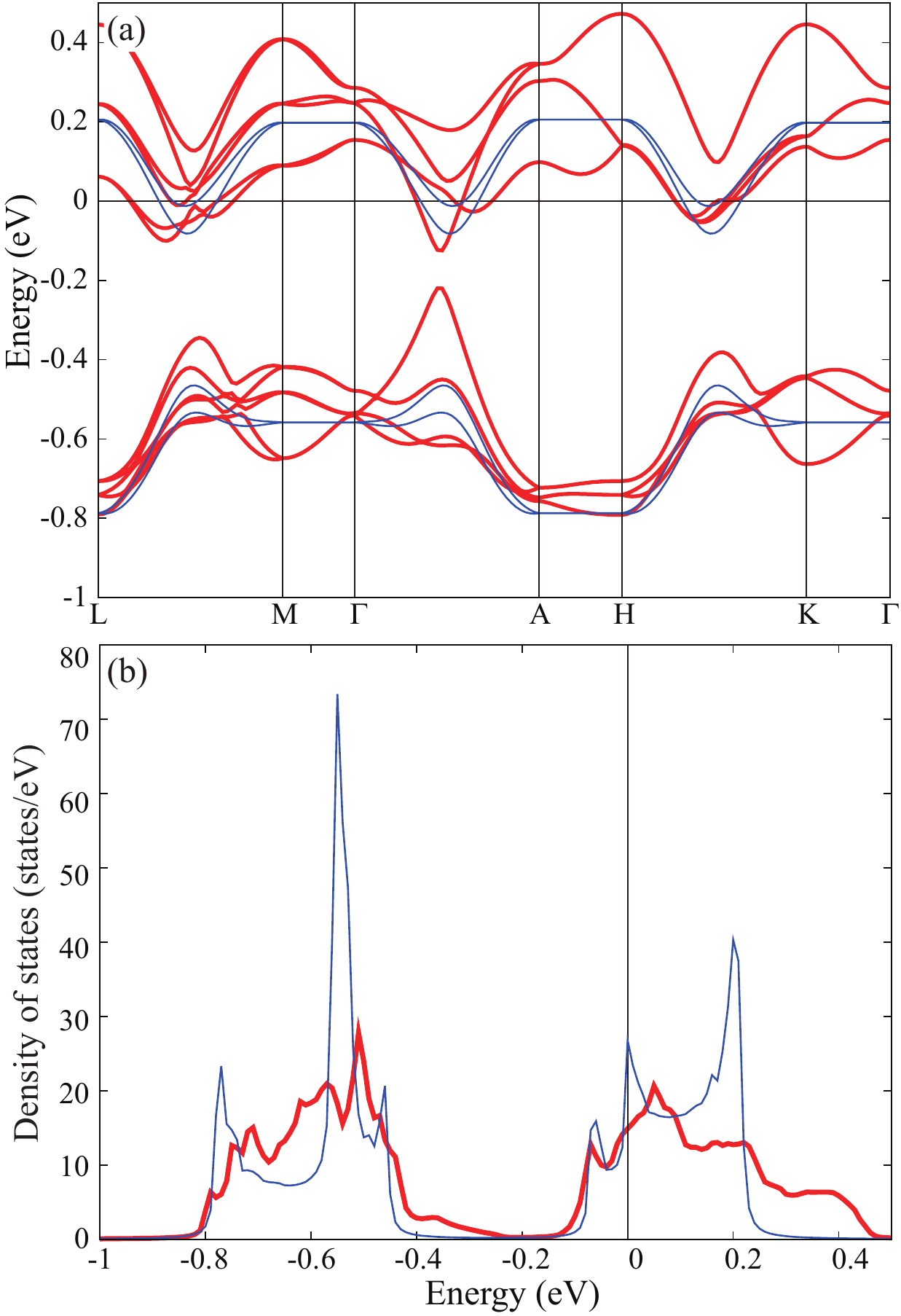}
\end{center} 
\vspace{-0.5cm} 
\caption{Effects of interchain electron transfers on the band structure: (a) Band dispersion and (b) density of states (DOS). The thick-red and thin-blue curves are the results with and without the interchain electron transfers, respectively. The DOS calculation is based on 37$\times$37$\times$37-$k$ point sampling, and the broadening of $\delta=0.01$~eV is applied. Also, in these calculations, we recalculated the Fermi energy to preserve the electron filling.}  
\label{band_with_without_interchain_transfer}
\end{figure}

We next remark the $IC$ electron transfer further in detail. This transfer is as small as 0.032~eV at maximum, which confirms the quasi-one-dimensional character of electrons. However, it is important to describe the details of the effect of the $IC$ transfer on the band structure of the $d_{xy}/d_{yz}$ Hamiltonian. Figure~\ref{band_with_without_interchain_transfer} shows the effect of the $IC$ electron transfer on the band structure. The panels (a) and (b) are the calculated band dispersion and density of states, respectively. The thick-red and thin-blue curves are the results with and without the $IC$ electron transfers, respectively. We found that it is important to include the $IC$ electron transfers larger than 0.003~eV in in order to quantitatively reproduce the original band structure. In the absence of $IC$ electron transfers, a fairly large gap of about 0.4~eV is generated due to the spin-orbit interaction, and we see separated upper and lower bands. Switching on the $IC$ electron transfer, the bandwidth of each band is widened. The $IC$ electron transfer effect is appreciable for the bands along the $\Gamma$-$A$ line.

\subsection{Interaction parameters}
Interaction parameters in the $d_{xy}/d_{yz}$ Hamiltonian are evaluated with cRPA. In cRPA, the constrained polarization function is first evaluated by switching off the transitions between specific occupied and unoccupied bands; since we want to derive the effective interaction parameters of the $d_{xy}/d_{yz}$ Hamiltonian, we exclude the band pairs involving Ir $d_{xy}$ and $d_{yz}$ orbitals in the polarization calculation~\cite{cRPA_Nakamura_2008}. The effective interaction $W({\bf r}, {\bf r}^{\prime}, \omega)$ is then evaluated using the resulting cRPA polarization function. Finally, we calculate the matrix elements of the static $W({\bf r}, {\bf r}^{\prime}, 0)$ with the $d_{xy}$ and $d_{yz}$ maximally localized Wannier functions, which gives $U_{i {\bf R} j {\bf R}'}^{\sigma\rho}$ [Eq.~(\ref{Wij})], $J_{i {\bf R} j {\bf R}'}^{\sigma\rho}$ [Eq.~(\ref{Jij})], and $K_{i {\bf R} j {\bf R}'}^{\sigma\rho}$ [Eq.~(\ref{Kij})]. 

TABLE~\ref{interaction} shows our derived interaction parameters of Ca$_5$Ir$_3$O$_{12}$. The low energy interactions are given by the cRPA values; RPA and bare interactions are provided only for comparison and discussion. In xTAPP band calculation with the valence-type pseudopotential, onsite cRPA intra-orbital interaction $U$ is estimated as 2.41~eV, where $U$ is evaluated as $U_{xy{\bf 0},xy{\bf 0}}^{\uparrow\downarrow}$. We note that $U_{xy{\bf 0},xy{\bf 0}}^{\uparrow\downarrow}=U_{xy{\bf 0},xy{\bf 0}}^{\downarrow\uparrow}=U_{yz{\bf 0},yz{\bf 0}}^{\uparrow\downarrow}=U_{yz{\bf 0},yz{\bf 0}}^{\downarrow\uparrow}$ is satisfied. Onsite cRPA inter-orbital interaction $U'$ is 1.93~eV, which is evaluated as $U_{xy{\bf 0},yz{\bf 0}}^{\uparrow\uparrow}$. Note that there are many symmetrically equivalent interactions that give the same value. There are three onsite exchange integrals, which are characterized by $J_{xy{\bf 0},yz{\bf 0}}^{\uparrow\uparrow}$, $J_{xy{\bf 0},yz{\bf 0}}^{\uparrow\downarrow}$, and $K_{xy{\bf 0},yz{\bf 0}}^{\uparrow\uparrow}$. As abbreviation, we simply write them $J^{\uparrow\uparrow}$, $J^{\uparrow\downarrow}$, and $K^{\uparrow\downarrow}$, which are corresponding to the Hund-type, exchange-type, and pair-hopping-type exchange integrals, respectively (Appendix~\ref{app:J-and-K}). These cRPA values are nearly 0.21~eV. Now, the obtained  $J$ reasonably satisfies the relation $U=U'+2J$ for the spherical atom. We note that the obtained cRPA values of $U \sim 2.4$~eV and $U' \sim 2$~eV look reasonable in terms of the previous {\it ab initio} estimates of $U$ and $U'$ for the $t_{2g}$ electrons in other Ir compounds (Sr$_2$IrO$_4$~\cite{Arita_2012} and Na$_2$IrO$_3$~\cite{Yamaji_2014}). 

The orbital-averaged $NN$ interaction $V_{NN}$ is estimated as 0.96~eV. Since the averaged $NN$ transfer $t=(t_{xy\uparrow,yz\uparrow}^{NN}+t_{yz\uparrow,xy\uparrow}^{NN})/2$ is about 0.197~eV, the correlation degree of freedom $(U-V_{NN})/t$ is estimated as 7.36. Then, in this compound, the electronic correlation is expected to play a substantial role in the low-energy physics. We note that the SOI estimated as 0.213~eV is comparable to the exchange interaction $J$ ($\sim$ 0.21~eV) and the largest $NN$ transfer (=0.215~eV). Therefore, the Hund physics and spin-orbit physics compete with each other and participate in the low-energy physics. We note in passing that the orbital-averaged nearest $IC$ interaction $V_{IC}$ is 0.51~eV.

The screening trend seen from the resulting bare, cRPA, and RPA interactions is normal~\cite{KazumaNakamura_2021}; the screening effect significantly affects the direct Coulomb integrals ($U$, $U^{\prime}$, $V_{NN}$, and $V_{IC}$)~
and is not so significant in the exchange integrals ($J^{\uparrow\uparrow}$, $J^{\uparrow\downarrow}$, and $K^{\uparrow\downarrow}$).

TABLE~\ref{interaction} also compares the differences in the calculation results between the semicore-type and valence-type pseudopotential for Ir. Basically, there is no substantial difference (interaction terms for both calculations differ by no more than a few percent). As far as the comparison in the results based on xTAPP is concerned, the resulting onsite cRPA Coulomb interaction $U$ with the semicore-type pseudopotential is 2.48~eV which is slightly larger than the value of 2.41~eV with the valence-type pseudopotential. This small increase is also observed for all other cRPA interaction terms, but also, interestingly, for all bare interaction terms: for instance, the on-site bare interaction increases from $9.83$~eV (for the valence-type pseudopotential) to $10.00$~eV (for the semicore-type pseudopotential).
We interpret the origin of this increase as the lower values of the cutoff radii for the semicore-type pseudopotential, with respect to the valence-type pseudopotential (as shown in Sec. \ref{sec_implcalc}), which slightly increases the localization of Wannier functions. Finally, the estimated correlation strength $(U-V_{NN})/t$ with the semicore-type pseudopotential is 7.36 with $t$ =0.201~eV,  
where the nearest-neighbor ($NN$) transfers with the semicore-type pseudopotential $t_{xy\uparrow,yz\uparrow}^{NN}$ and $t_{yz\uparrow,xy\uparrow}^{NN})$ are 0.182 and 0.219~eV, respectively.

On the other hand, when compared with the results based on the {\sc Quantum Espresso} calculations, which are also based on the semicore-type pseudopotential for Ir, the interaction values are almost the same as the results of xTAPP; in cRPA, $U$ = 2.47~eV for {\sc Quantum Espresso} and $U$ = 2.48~eV for xTAPP-semicore-pseudopotential. Thus, in {\sc Quantum Espresso}, since the averaged $NN$ transfer $t$ and interaction $V$ is about 0.201~eV and 1.01~eV, respectively, the correlation strength $(U-V)/t$ is estimated as 7.28 (which is close to the value 7.36 for xTAPP). Then, we evaluate that the correlation strength of Ca$_5$Ir$_3$O$_{12}$ is about 7.3.  
\begin{table*}[htpb] 
\caption{Calculated static interaction parameters of the $d_{xy}/d_{yz}$ Hamiltonian of Ca$_5$Ir$_3$O$_{12}$. The interaction parameters with the bare (unscreened), constrained RPA (cRPA), and usual RPA are compared. $U$ and $U^{\prime}$ are onsite intra-orbital and inter-orbital direct Coulomb integrals, respectively. Also, $J^{\uparrow\uparrow}$, $J^{\uparrow\downarrow}$, and $K^{\uparrow\downarrow}$ are onsite exchange integrals, which are corresponding to the Hund-type, exchange-type, and pair-hopping-type exchange integrals, respectively (Appendix~\ref{app:J-and-K}). $V_{NN}=(1/N_w)^2\sum_{i,j=1}^{N_w} U_{i{\bf 0}j{\bf R}}^{\uparrow\uparrow}$ is the orbital-averaged value of the nearest-neighbor ($NN$) interactions with ${\bf R}=(0,0,1)$ and $N_w=2$, where $i$ and $j$ specify the $xy$ and $yz$ orbitals. Also, $V_{IC}$ is the orbital-averaged value of the nearest interchain ($IC$) interactions. The table also compares the results based on the two band-calculation software (xTAPP and {\sc Quantum Espresso}). The ``(s)" and ``(v)" after SO-GGA and GGA indicate the pseudopotential types of Ir, i.e., the semicore-type or valence-type. The unit of the interaction parameter is eV.} 
\begin{center} 
\begin{tabular}{l@{\ \ \ }c@{\ \ \ }c@{\ \ \ }c@{\ \ \ }c@{\ \ \ }
c@{\ \ \ }c@{\ \ \ }c@{\ \ \ }c@{\ \ \ }c@{\ \ \ }c@{\ \ \ }c@{\ \ \ }c@{\ \ \ }
c@{\ \ \ }c@{\ \ \ }c@{\ \ \ }c@{\ \ \ }c@{\ \ \ }c@{\ \ \ }c@{\ \ \ }c} 
 \hline \hline \\ [-5pt]
 & \multicolumn{11}{c}{xTAPP}  & & \multicolumn{7}{c}{{\sc Quantum Espresso}} \\ 
 \cmidrule{2-12} \cmidrule{14-20} \\ 
                             & \multicolumn{3}{c}{SO-GGA (s)}     &  & \multicolumn{3}{c}{SO-GGA (v)}   & & \multicolumn{3}{c}{GGA (v)}        & & \multicolumn{3}{c}{SO-GGA (s)}& & \multicolumn{3}{c}{GGA (s)}\\ 
 \cmidrule{2-4} \cmidrule{6-8} \cmidrule{10-12} \cmidrule{14-16} \cmidrule{18-20} 
                             & bare  & cRPA & RPA  & & bare & cRPA & RPA  & & bare  & cRPA & RPA  & & bare & cRPA & RPA  & & bare  & cRPA & RPA 
 \\ \hline \\ [-5pt] 
   $U$                       & 10.00 & 2.48 & 0.42 & & 9.83 & 2.41 & 0.41 & & 10.01 & 2.29 & 0.40 & & 9.98 & 2.47 & 0.42 & & 10.15 & 2.44 & 0.40 \\
   $U'$                      & 9.32  & 1.98 & 0.13 & & 9.15 & 1.93 & 0.13 & & 9.26  & 1.74 & 0.09 & & 9.29 & 1.98 & 0.13 & & 9.38  & 1.87 & 0.09 \\
   $J^{\uparrow\uparrow}$    & 0.28  & 0.23 & 0.14 & & 0.27 & 0.21 & 0.14 & & 0.29  & 0.23 & 0.14 & & 0.28 & 0.23 & 0.14 & & 0.30  & 0.25 & 0.14 \\
   $J^{\uparrow\downarrow}$  & 0.26  & 0.21 & 0.13 & & 0.25 & 0.20 & 0.12 & & 0.29  & 0.23 & 0.14 & & 0.26 & 0.21 & 0.13 & & 0.30  & 0.25 & 0.14 \\
   $K^{\uparrow\downarrow}$  & 0.28  & 0.23 & 0.14 & & 0.27 & 0.21 & 0.14 & & 0.29  & 0.23 & 0.14 & & 0.28 & 0.23 & 0.14 & & 0.30  & 0.25 & 0.14 \\
   $V_{NN}$                  & 4.44  & 1.00 & 0.04 & & 4.42 & 0.96 & 0.04 & & 4.39  & 0.81 & 0.05 & & 4.47 & 1.01 & 0.05 & & 4.44  & 0.91 & 0.06 \\
   $V_{{IC}}$                & 2.86  & 0.54 & 0.00 & & 2.85 & 0.51 & 0.00 & & 2.86  & 0.46 & 0.00 & & 2.86 & 0.54 & 0.00 & & 2.87  & 0.51 & 0.00 \\
%
%

\hline \hline
\end{tabular} 
\end{center} 
\label{interaction} 
\end{table*} 

TABLE~\ref{interaction-t2g} compares the interaction parameters of the $t_{2g}$ and $d_{xy}/d_{yz}$ Hamiltonian, where the $t_{2g}$ Hamiltonian consider not only $d_{xy}$ and $d_{yz}$ orbitals but also the $d_{zx}$ orbital. As a result, the derived interaction parameters of the $t_{2g}$ Hamiltonian become larger than those of the $d_{xy}/d_{yz}$ Hamiltonian, because the polarizations taking place within all the $t_{2g}$ electrons are excluded in cRPA and the screening gets smaller than in the $d_{xy}/d_{yz}$ case. As seen from this result, the direct Coulomb integrals $U$ and $U^{\prime}$ of the $t_{2g}$ Hamiltonian is about 0.3~eV larger than those of the $d_{xy}/d_{yz}$ Hamiltonian. Also, the difference in the exchange integrals between the $t_{2g}$ and $d_{xy}/d_{yz}$ Hamiltonian is as small as 0.01~eV. We note that our estimated $t_{2g}$-$U$ of Ca$_5$Ir$_3$O$_{12}$ ($\sim$ 2.79~eV as an average of $U_{xy}$, $U_{yz}$, and $U_{zx}$) is close to the $U$ value of Na$_2$IrO$_3$ (2.72~eV)~\cite{Yamaji_2014}. On the other hand, the orbital averaged $t_{2g}$-$U$ value of Sr$_2$IrO$_4$~\cite{Arita_2012} has been reported as about 2.26~eV. The systematic study of material dependence of the degree of strong electronic correlation for the Ir oxides will be a  very interesting issue, which is left for future problems. 

TABLE~\ref{interaction-t2g} also gives a comparison with the interaction parameters obtained with the Wannier functions without the Wannier spread minimization, that is, the Wannier functions obtained just after the initial-guess projection. This model is denoted as $d_{xy}^{ini}/d_{yz}^{ini}$. We see from the result that the derived interaction parameters of the $d_{xy}/d_{yz}$ and $d_{xy}^{ini}/d_{yz}^{ini}$ Hamiltonians are almost the same. Therefore, in the parameter derivation, the quantitative effect of the spread-functional minimization on the interaction parameter is small. 
\begin{table}[htpb] 
\caption{Comparison of calculated static constrained RPA (cRPA) interaction parameters between the two-orbital $d_{xy}/d_{yz}$ and three-orbital $t_{2g}$ Hamiltonians. The bare (unscreened) interaction values are also listed for reference. The definition of the parameters is the same as TABLE~\ref{interaction}. The $d_{xy}^{ini}/d_{yz}^{ini}$ column lists calculated interaction parameters with the initial-guess Wannier functions (i.e., results obtained without the spread functional minimization). The unit of the interaction integral is eV.}   
\begin{center} 
\begin{tabular}{l@{\ \ \ }c@{\ \ \ }c@{\ \ \ }c@{\ \ \ }c@{\ \ \ }c@{\ \ \ }c@{\ \ \ }c@{\ \ \ }c} 
 \hline \hline \\ [-5pt]
  & \multicolumn{2}{c}{$t_{2g}$} & & \multicolumn{2}{c}{$d_{xy}/d_{yz}$} & & \multicolumn{2}{c}{$d_{xy}^{ini}/d_{yz}^{ini}$} \\ 
    \cmidrule{2-3}                \cmidrule{5-6}             \cmidrule{8-9} 
                                   & bare   & cRPA  & & bare  & cRPA  & & bare  & cRPA  \\ 
 \hline \\ [-5pt] 
 $U_{xy}$                          & 10.044 & 2.737 & & 9.834 & 2.411 & & 9.831 & 2.410 \\ [+5pt] 
 $U_{zx}$                          & 10.588 & 2.881 & & -     & -     & & -     & -     \\ [+5pt] 
 $U_{xy,yz}^{\prime}$              & 9.286  & 2.218 & & 9.155 & 1.926 & & 9.151 & 1.925 \\ [+5pt] 
 $U_{xy,zx}^{\prime}$              & 9.229  & 2.103 & & -     & -     & & -     & -     \\ [+5pt] 
 $J_{xy,yz}^{\uparrow\uparrow}$    & 0.285  & 0.227 & & 0.270 & 0.212 & & 0.270 & 0.212 \\ [+5pt] 
 $J_{xy,yz}^{\uparrow\downarrow}$  & 0.283  & 0.225 & & 0.251 & 0.198 & & 0.252 & 0.198 \\ [+5pt] 
 $K_{xy,yz}^{\uparrow\downarrow}$  & 0.285  & 0.227 & & 0.270 & 0.212 & & 0.270 & 0.212 \\ [+5pt] 
 $J_{xy,zx}^{\uparrow\uparrow}$    & 0.285  & 0.229 & & -     & -     & & -     & -                         \\ [+5pt] 
 $J_{xy,zx}^{\uparrow\downarrow}$  & 0.280  & 0.226 & & -     & -     & & -     & -                         \\ [+5pt]
 $K_{xy,zx}^{\uparrow\downarrow}$  & 0.285  & 0.229 & & -     & -     & & -     & -                         \\ [+5pt] 
 $V_{{\rm NN}}$                    & 4.523  & 1.119 & & 4.425 & 0.964 & & 4.423 & 0.963 \\ [+5pt] 
 $V_{{\rm IC}}$                    & 2.866  & 0.597 & & 2.855 & 0.514 & & 2.854 & 0.514 \\ 
 %
 %
 %
 [+5pt] \hline \hline
\end{tabular} 
\end{center} 
\label{interaction-t2g} 
\end{table}

\section{Summary and discussion}\label{sec_summary_discussion} 
In summary, we presented an {\it ab initio} framework to study the effective Hamiltonian for strongly correlated electron systems with strong SOI; the spinor formalism and algorithm for deriving the effective Hamiltonian represented in the maximally localized Wannier function are presented and implemented in open source software RESPACK. In particular, we described how to use crystal symmetries of the material in the presence of SOI; (i) a proper initial guess setting for the Wannier spinor involving proper choices of the quantization axis, and (ii) the computational procedure for generating all the $k$-point wave function data from the irreducible $k$-point data, which are useful for the speedup of calculations and memory savings.  

As an example of application, we derived an {\it ab initio} effective low-energy Hamiltonian of Ca$_5$Ir$_3$O$_{12}$. Ir $t_{2g}$ electrons are described with the Wannier spinors represented in the local coordinate system fixed to the IrO$_6$ octahedron. After analyzing the band structure, we found that the effective Hamiltonian described in terms of the degenerate $d_{xy}$ and $d_{yz}$ orbitals offers the following interesting insights into characters and properties of this compound:
    \begin{figure*}[t!] 
     \begin{center} 
     \includegraphics[width=0.79\textwidth]{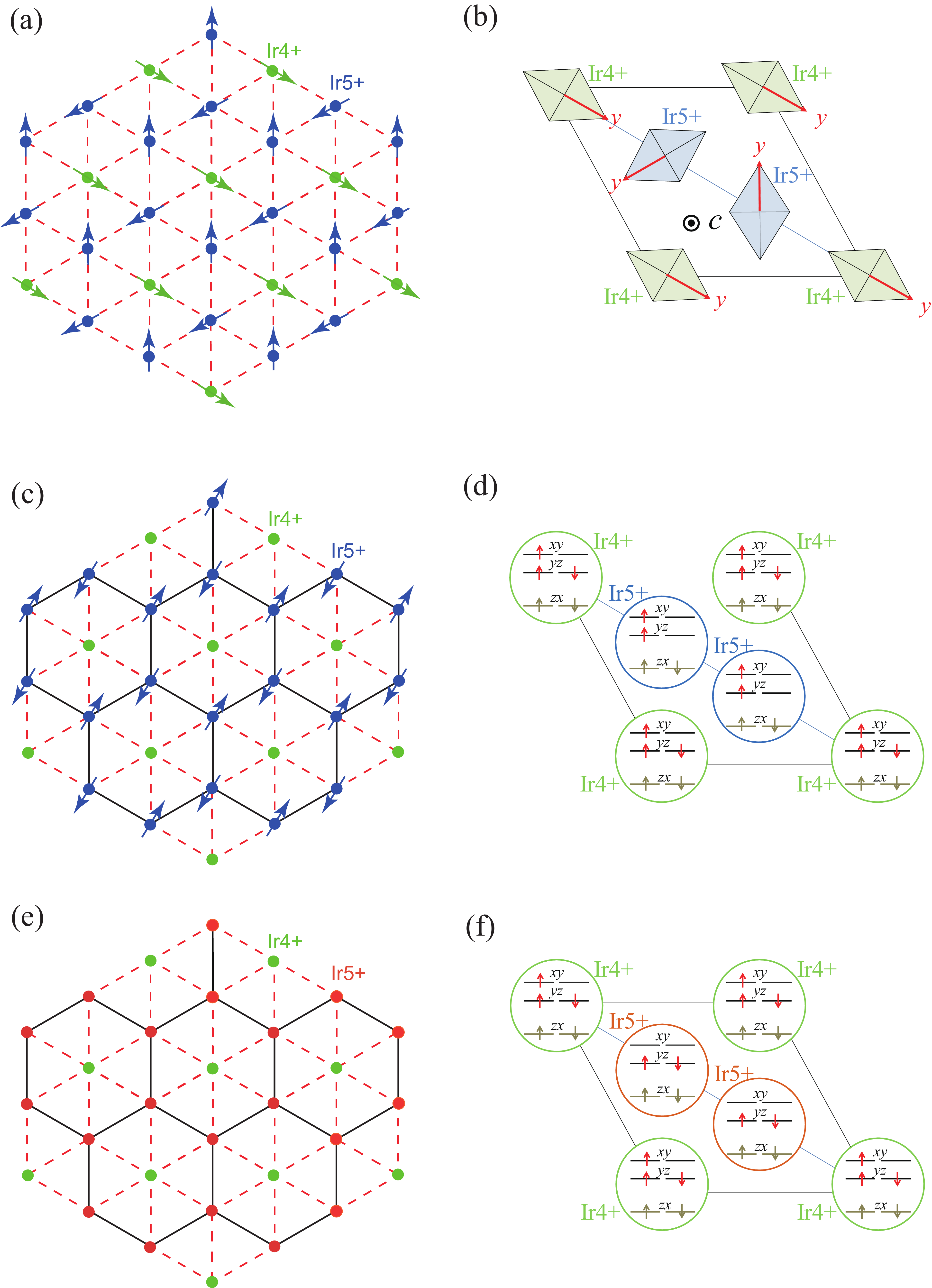}
     \end{center} 
     \vspace{-0.5cm} 
     \caption{Possible spin-charge orders to dissolve the frustration, based on the derived parameters for the $d_{xy}/d_{yz}$ Hamiltonian: (a) The charge  pattern including ferrimagnetic 120$^{\circ}$ spin structure, where Ir$^{4+}$ and Ir$^{5+}$ sites are assumed to be spin-1/2 and spin-1, respectively, and described by green and blue dots. In this figure, the Ir spin is along the local $y$ axis of the IrO$_6$ octahedron [panel (b)], because the magnetic easy axis is the local $y$ axis (Appendix~\ref{app:SingleChainHamiltonian}). The panel (c) shows a possible charge order and accompanied partial magnetic order at the hexagonal Ir$^{5+}$ (spin-1) sublattice, which can generate strong quantum fluctuation on the triangular Ir$^{4+}$ (spin-1/2) sublattice. The spin patterns of the panel (a) and (c) are based on the Ir electronic configurations with the Hund interaction $>$ spin-orbit interaction (SOI) [panel (d)]. Another possible spin-charge pattern drawn in the panel (e) exists, where the Hund interaction is less than SOI. In this case, Ir$^{5+}$ sites denoted by orange dots are assumed to be spin-0 [panel (f)]. In the panels (c) and (e), the unpaired spins on the Ir$^{4+}$ sites are expected to dynamically fluctuate and hardly order because of the vanishing mean field from the surrounded Ir$^{5+}$ electrons.} 
     \label{Spin-Charge_Order}
    \end{figure*}
\begin{enumerate}
    \item 
    The electronic correlation strength characterized by $(U-V)/t$ is estimated as $\sim$ 7, where $t$ is the averaged transfer of the nearest pair, and $U$ and $V$ are onsite and nearest interactions, respectively (see TABLEs~\ref{transfer} and \ref{interaction}). A large electron correlation is expected to play a key role in the low-energy physics.
    \item 
    The estimated value of the largest transfer taking account of SOI is 0.21~eV, which is as large as the exchange interaction 0.20-0.23~eV. This amplitude is, however, relatively smaller than the crystal field splitting between the $d_{yz}/d_{zx}$ orbitals and $d_{zx}$ orbital ($\sim 1$~eV). This makes the dominant role of the crystal field splitting in the $t_{2g}$ electron levels over the spin-orbit splitting in contrast to Sr$_2$IrO$_4$~\cite{Arita_2012}, where the bands are rather characterized by the total angular momentum $J_{\rm eff}$ that is a good quantum number in the strong SOI limit. 
    \item 
    Although one-dimensional anisotropy exists, it is important to include the $IC$ electronic transfer equal to or larger than 0.003~eV to reproduce the overall quantitative band structure, where the dispersion in the $IC$ direction is not negligible (see Fig.~\ref{band_with_without_interchain_transfer}).
    \item
    In the present case, the SOI effect plays a crucial role by forming a gap below the Fermi level within the $d_{xy}/d_{yz}$ manifold. However, the partially 1/6 filled upper bands constituted by $d_{xy}/d_{yz}$ orbitals, separated by the spin-orbit gap from the lower $d_{xy}/d_{yz}$ bands, is expected to generate the Fermi surface and metallic conduction, where the $IC$ electron transfer is not negligible and the one-dimensional localization effect should be limited (see the item~3 above). The bad metallic behavior may thus be ascribed to the electronic correlation presumably combined with the valence fluctuation.
    \item
    Since the effective onsite Coulomb repulsion ($>2$~eV) is larger than the spin-orbit gap and furthermore larger than the total $d_{yz}/d_{xy}$ band width ($\sim$ 1.5~eV), the final electronic structure could be totally reconstructed by the electronic correlation, which makes it necessary to accurately solve the effective Hamiltonian derived here beyond the DFT-GGA level. Even the explicit inclusion of the $d_{zx}$ orbital into the effective Hamiltonian might be necessary to consider because of the large onsite Coulomb repulsion $U_{zx}$ (see TABLE~\ref{interaction-t2g}). 
    \item
    The geometrical frustration arising from the triangular Ir configuration may be dissolved by a regular alignment of two Ir$^{5+}$ and one Ir$^{4+}$ on each triangle, namely charge ordering as is illustrated in Fig.~\ref{Spin-Charge_Order}. We give in the panel (a) a possible charge and spin arrangement in the $ab$ plane, where this configuration will minimize the loss of intersite Coulomb energy. For clear understanding of the spin pattern accompanied with the charge arrangement, we depict in the panel (b) the orientation of IrO$_6$ octahedron and the local $y$ axis defined for each octahedron. We note that the local $y$-axis is the easy axis of an Ir spin (see the item 7 below). Then, a naive expectation is 120$^{\circ}$ spin structure consisting of one spin-1/2 Ir$^{4+}$ and two spin-1 Ir$^{5+}$, where the Ir spin drawn in Fig.~\ref{Spin-Charge_Order} (a) is along the local-$y$ axis. 
    \item
    In the energy or temperature scale lower than the SOI, the magnetic anisotropy induced by the SOI will play an important role to stabilize discrete broken symmetry phases: As explained below, the SOI induces an easy-axis anisotropy along the local $y$ axis. The geometry and crystal field splitting of the local distorted octahedral cluster IrO$_6$ essentially determine the magnetic anisotropy. As shown in the onsite energy diagram (Fig. \ref{onsite-energy}), when the SOI is neglected, the $d_{xy}$ and $d_{yz}$ orbitals are nearly degenerated, thus, may induce an angular momentum along the local $y$ axis. Then, the atomic SOI is dominated by couplings between the $y$ components of the spin and angular momentum, $\lambda S^y L^y$, where $\lambda$ is the effective spin-orbit coupling constant and $S^y$ ($L^y$) is the $y$ component of the spin (angular momentum) of the Ir ions in the local coordinate. The spin-orbit coupling $\lambda S^y L^y$ may stabilize the total angular momentum along the local $y$ axis in the broken symmetry phases at low temperatures [Fig.~\ref{Spin-Charge_Order} (a)]. In Appendix \ref{app:SingleChainHamiltonian}, the relevance of $\lambda S^y L^y$ is demonstrated by using the transfer integrals of the $d_{xy}$/$d_{yz}$ Hamiltonian (given in TABLE~\ref{transfer} and Fig.~\ref{trasfer_definition}).
    \item 
    An alternative expectation is the hexagonal lattice of Ir$^{5+}$ atoms with the center of hexagon occupied by spin-1/2 Ir$^{4+}$, which is shown in Fig.~\ref{Spin-Charge_Order} (c). The antiferromagnetic intra- and inter-chain exchange interactions suggested by the high-temperature magnetic susceptibility measurement is expected to be larger between two Ir$^{5+}$ atoms than the exchange involving spin-1/2 Ir$^{4+}$ atom, because of the spin-1 Ir$^{5+}$ state due to the Hund's rule coupling. This picture is based on the electronic configuration of the Ir atoms, described in Fig.~\ref{Spin-Charge_Order} (d). In this spin structure, partial antiferromagnetic order of the hexagonal Ir$^{5+}$ atoms would be formed and leave the spin-1/2 Ir$^{4+}$ sites disordered because of the cancellation of the antiferromagnetic coupling from the neighboring Ir$^{5+}$ spins. In the chain direction, there may exist a competition between the two possible cases: To reduce the intrachain and intersite Coulomb repulsion, the Ir$^{4+}$ sites are aligned in every three sites within a chain and alternatingly between chains as well to minimize the intersite interaction. On the other hand, the kinetic energy is lowered by forming a chain with uniform Ir$^{4+}$ valence and two uniform Ir$^{5+}$ chains. The antiferromagnetic order becomes stronger in the chain direction in the latter case and lowers the energy as well. In the lower energy (temperature) scale, the 120 degree coplanar order of the moment on the triangular Ir$^{4+}$ sublattice or the spin alignment perpendicular to the Ir$^{5+}$ moment represented by $e^{+i\theta /2}|\uparrow\rangle \pm e^{-i\theta/2}|\downarrow\rangle$ in the basis of the Ir$^{5+}$ moment axis, where $\theta$ is a real constant, may also subsequently occur, which may further cause the ferrimagnetism perpendicular to the Ir$^{5+}$ moment in the latter case. 
    \item 
    Another possibility is that the SOI is dominant beyond the Hund's coupling. In this case, Ir$^{5+}$ sites become spin-0 atoms, which is shown in Figs.~\ref{Spin-Charge_Order} (e) and (f). In the present estimation, the SOI $\sim$ 0.21~eV compete with the exchange integral $\sim$ 0.22~eV. So, various spin and charge orders including Figs~\ref{Spin-Charge_Order} (a), (c), and (e) will compete at low energies. In general, in a situation where the spin-orbit interaction is large, the multipole degree of freedom may become apparent. These issues will also need to be considered carefully.   
    \item
    Although such a spin-charge-order correlation might exist as a short-range fluctuation, the preserved original symmetry at low temperatures so far experimentally reported suggests that the Ir$^{4+}$ and Ir$^{5+}$ configurations are not regular and static but dynamic or random at least above 15 K. It is desired to see the crystal symmetry, charge and spin order/fluctuation in the low-temperature phase below 7.8 K, whether the spin-charge order or glassy freezing may take place together with semiconducting behaviors. It is also desired to specify the time scale of the valence fluctuation in experimental measurements and dependence of spin-charge correlation on the annealing rate to reach the low-temperature phase. The nonlinear conduction \cite{Matsuhira_2018} may also be originated from such fluctuations and weak pinning by the disorder, which could induce low-frequency noise in the AC transport response as well.
\end{enumerate}

Because of the above interplay among the SOI, electronic correlation, valence fluctuation and geometrical frustration in the magnetic coupling Ca$_5$Ir$_3$O$_{12}$ provides us with an intriguing playground of competition and fluctuation to be elucidated in the future by solving the effective Hamiltonian derived here and by comparing with refined experiments.

\section{Acknowledgments}
\label{acknowledgments}
This work was supported by Fonds de recherche du Qu\'{e}bec - Nature et technologies (FRQNT). We also acknowledge the financial support of JSPS Kakenhi Grant No. 16H06345 (MC, JP, KN, YN, TT, YY, YY, and MI), No. 17K14336 (YN), No. 18H01158 (YN), No. 20K14423 (YN), No. 16K05452 (KN), No. 17H03393 (KN), No. 17H03379 (KN), and No. 19K03673 (KN). 
The authors are grateful to the MEXT HPCI Strategic Programs, and the Creation of New Functional Devices and High-Performance Materials to Support Next Generation Industries (CDMSI) for their financial support. This work was supported by MEXT as "Program for Promoting Researches on the Supercomputer Fugaku" (Basic Science for Emergence and Functionality in Quantum Matter). We also acknowledge the support provided by the RIKEN Advanced Institute for Computational Science under the HPCI System Research project (Grants No. hp180170, hp190145 and hp200132). A part of the computation was done at Supercomputer Center, Institute for Solid State Physics, University of Tokyo. Also, another part of the computation was done by using the supercomputer system or the intercloud system or Hokkaido University High-Performance Intercloud at the information initiative center, Hokkaido University, Sapporo, Japan.

\newpage 

\appendix
\begin{widetext}
\section{Derivation of the formula for the effective Hamiltonian} 
\label{app:Hderivation-KN}
In this appendix, we consider a derivation of the effective Hamiltonian in Eq.~(\ref{eq:H}). 
We start from the general form of Hamiltonian described in terms of the spinor field operator as 
\begin{eqnarray}
    \hat{H} 
    &=& \int_{V} d{\bf r} 
    \hat{\Phi}^{\dagger}({\bf r})
    {\cal H}_0({\bf r})
    \hat{\Phi}({\bf r})  
    +\frac{1}{2}  
    \int_{V} \int_{V} d{\bf r} d{\bf r}'
    \hat{\Phi}^{\dagger}({\bf r})
    \hat{\Phi}^{\dagger}({\bf r}')
    {\cal W}({\bf r},{\bf r}',\omega=0) 
    \hat{\Phi}^{}({\bf r}')
    \hat{\Phi}^{}({\bf r}), 
\label{eq:Hbegin-KN}    
\end{eqnarray}
\end{widetext}
where $\hat{\Phi}({\bf r})$ and  $\hat{\Phi}^{\dagger}({\bf r})$ are the spinor field operators. 
${\cal H}_0$ is a one-body Hamiltonian which is a $2\times2$ matrix. ${\cal W}$ is the static limit of the frequency-dependent effective interaction, which is assumed to be a scalar form without spin dependence~\cite{Aryasetiawan2008,Aryasetiawan2009}. The integrals in Eq.~(\ref{eq:Hbegin-KN}) are over the crystal volume $V$. For ${\cal H}_0$, we use the Kohn-Sham Hamiltonian, and, for ${\cal W}({\bf r}, {\bf r}^{\prime})$, we use the cRPA effective interaction;
\begin{eqnarray} 
{\cal H}_0({\bf r}) &\sim& {\cal H}_{{\rm KS}}({\bf r}), \label{HKS-KN} \\
{\cal W}({\bf r}, {\bf r}^{\prime}) &\sim& W({\bf r}, {\bf r}^{\prime}). \label{W-KN}  
\end{eqnarray} 

The spinor field operators are expressed with the Wannier spinor function $\Phi_{i\sigma{\bf R}}({\bf r})$ and $\Phi^{\dagger}_{i\sigma{\bf R}}({\bf r})$ defined in Eqs.~(\ref{spinor-wannier}) and (\ref{spinor-wannier-d}) and their creation/annihilation operators $a_{i\sigma{\bf R}}^{\dagger}$ and $a_{i\sigma{\bf R}}$ as 
\begin{eqnarray}
    \hat{\Phi}({\bf r}) 
    = \sum_{i\sigma} \sum_{\bf R} 
    \Phi_{i\sigma{\bf R}}({\bf r}) 
    a^{}_{i\sigma{\bf R}}, \label{Phi-hat-KN}
\end{eqnarray}
and 
\begin{eqnarray} 
    \hat{\Phi}^{\dagger}({\bf r}) 
    &=& \sum_{i\sigma} \sum_{\bf R} 
    \Phi^{\dagger}_{i\sigma{\bf R}}({\bf r}) 
    a^{\dagger}_{i\sigma{\bf R}}. \label{Phi-hat-d-KN} 
\end{eqnarray}
By inserting Eqs.~(\ref{HKS-KN}), (\ref{W-KN}), (\ref{Phi-hat-KN}), and (\ref{Phi-hat-d-KN}) into Eq.~(\ref{eq:Hbegin-KN}), we obtain 
\begin{widetext}
\begin{eqnarray}
    \hat{H} 
    &=& 
    \sum_{i\sigma, j\rho} 
    \sum_{{\bf R}_i, {\bf R}_j}   
    \int_V d{\bf r} 
    \Phi_{i\sigma{\bf R}_i}^{\dagger}({\bf r}) 
    {\cal H}_{{\rm KS}}({\bf r})
    \Phi_{j\rho{\bf R}_j}({\bf r}) 
    a^{\dagger}_{i\sigma{\bf R}_i}
    a^{}_{j\rho{\bf R}_j} \nonumber \\ 
    &+&
    \frac{1}{2}  
    \sum_{i\sigma, j\rho, k\lambda, l\nu}
    \sum_{{\bf R}_i,{\bf R}_j,{\bf R}_k,{\bf R}_l}
    \int_{V} \int_{V} d{\bf r} d{\bf r}^{\prime}
    \Phi^{\dagger}_{i\sigma{\bf R}_i}({\bf r})
    \Phi^{}_{l\nu{\bf R}_l}({\bf r})
    W_{}({\bf r},{\bf r}')
    \Phi^{\dagger}_{j\rho{\bf R}_j}({\bf r}')
    \Phi^{}_{k\lambda{\bf R}_k}({\bf r}')
    a^{\dagger}_{i\sigma{\bf R}_i} 
    a^{\dagger}_{j\rho{\bf R}_j}
    a^{}_{k\lambda{\bf R}_k} 
    a^{}_{l\nu{\bf R}_l}, 
\label{eq:Hbegincc3-KN}  
\end{eqnarray}
\end{widetext}
where $i, j, k$, and $l$ specify the Wannier orbital, and $\sigma$, $\rho$, $\lambda$, and $\nu$ are indices that specify the front and back components of the Kramers pair. ${\bf R}_i, {\bf R}_j, {\bf R}_k$, and ${\bf R}_l$ are indices for the lattice vector. 

For the inner product $\Phi^{\dagger}_{i\sigma{\bf R}_i}({\bf r})\Phi^{}_{l\nu{\bf R}_l}({\bf r})$ for the two-particle (namely, interaction) part proportional to $W$, we introduce ``colinear approximation''; we suppose that the product of the front components or that of the back components may have a significant value. Then, we obtain 
\begin{eqnarray}
 \Phi^{\dagger}_{i\sigma{\bf R}_i}({\bf r})
 \Phi^{}_{l\nu{\bf R}_l}({\bf r})
 \sim 
 \Phi^{\dagger}_{i\sigma{\bf R}_i}({\bf r}) 
 \Phi^{}_{l\sigma{\bf R}_l}({\bf r}) 
 \delta_{\sigma\nu}
 \label{inner-product-1} 
\end{eqnarray}
and 
\begin{eqnarray}
 \Phi^{\dagger}_{j\rho{\bf R}_j}({\bf r}^{\prime}) 
 \Phi^{}_{k\lambda{\bf R}_k}({\bf r}^{\prime})
 \sim 
 \Phi^{\dagger}_{j\rho{\bf R}_j}({\bf r}^{\prime})
 \Phi^{}_{k\rho{\bf R}_k}({\bf r}^{\prime})
 \delta_{\rho\lambda}. 
 \label{inner-product-2} 
\end{eqnarray}
For the Wannier functions sharing the same site, this approximation is exact, because, in the Wannier spinors forming the Kramers pair, the front-type Wannier spinor and the back-type Wannier spinor are exactly orthogonal to each other. On the other hand, in the case where the Wannier-spinor quantization axis is different for each site; i.e., non-colinear case, the front-type Wannier spinor and the distant back-type Wannier spinor are not orthogonal to each other. Then, the colinear approximation means dropping terms due to the product of the front-type and back-type Wannier spinors. However, since the spatial overlap between the 
distant Wannier functions is small, it does not seem to cause a serious quantitative error. 
\begin{widetext}
By inserting Eqs.~(\ref{inner-product-1}) and (\ref{inner-product-2}) into Eq.~(\ref{eq:Hbegincc3-KN}), we have 
\begin{eqnarray}
    \hat{H} 
    &=& 
    \sum_{i, j} \sum_{\sigma, \rho} 
    \sum_{{\bf R}_i, {\bf R}_j} 
    t_{i{\bf R}_i j{\bf R}_j}^{\sigma\rho} 
    a^{\dagger}_{i\sigma{\bf R}_i}
    a^{}_{j\rho{\bf R}_j} 
    +
    \frac{1}{2}  
    \sum_{i, j, k, l}
    \sum_{\sigma, \rho}
    \sum_{{\bf R}_i,{\bf R}_j,{\bf R}_k,{\bf R}_l}
    I_{i{\bf R}_i,j{\bf R}_j,l{\bf R}_l,k{\bf R}_k}^{\sigma\rho} 
    a^{\dagger}_{i\sigma{\bf R}_i} 
    a^{\dagger}_{j\rho{\bf R}_j}
    a^{}_{k\rho{\bf R}_k} 
    a^{}_{l\sigma{\bf R}_l} 
\label{eq:Hbegincc3}  
\end{eqnarray}
with the matrix elements of the one-particle part 
\begin{eqnarray}
t_{i{\bf R} j{\bf R}'}^{\sigma\rho}
=
\int_V d{\bf r} 
\Phi_{i\sigma{\bf R}}^{\dagger}({\bf r}) 
{\cal H}_{0}({\bf r})
\Phi_{j\rho{\bf R}'}({\bf r}) 
\end{eqnarray}
and the matrix elements of the two-particle part 
 \begin{eqnarray}
 I_{i{\bf R}_i,j{\bf R}_j,l{\bf R}_l,k{\bf R}_k}^{\sigma\rho} 
 = \int_{V} \int_{V} d{\bf r} d{\bf r}^{\prime}
    \Phi^{\dagger}_{i\sigma{\bf R}_i}({\bf r})
    \Phi^{}_{l\sigma{\bf R}_l}({\bf r})
    W_{}({\bf r},{\bf r}')
    \Phi^{\dagger}_{j\rho{\bf R}_j}({\bf r}^{\prime})
    \Phi^{}_{k\rho{\bf R}_k}({\bf r}^{\prime}). 
 \end{eqnarray}
\end{widetext}
Furthermore, the above four-center integral is approximated to the two-center integral; with $(i, {\bf R}_i)=(l, {\bf R}_l)$ and $(j, {\bf R}_j)=(k, {\bf R}_k)$, we obtain the direct Coulomb integral as    
\begin{eqnarray}
U_{i {\bf R} j {\bf R}'}^{\sigma\rho}  = 
I_{i{\bf R},j{\bf R'},i{\bf R},j{\bf R'}}^{\sigma\rho}, 
\end{eqnarray}
where we rewrite ${\bf R}_i$ and ${\bf R}_j$ as ${\bf R}$ and ${\bf R}^{\prime}$, respectively. Also, with the approximation of $(i, {\bf R}_i)=(k, {\bf R}_k)$ and $(l, {\bf R}_l)=(j, {\bf R}_j)$, the $(ij|ji)$-type exchange integral is obtained as  
\begin{eqnarray}
J_{i {\bf R} j {\bf R}'}^{\sigma\rho}  = 
I_{i{\bf R},j{\bf R'},j{\bf R'},i{\bf R}}^{\sigma\rho}.  
\end{eqnarray} 
Finally, with $(i, {\bf R}_i)=(j, {\bf R}_j)$ and $(l, {\bf R}_l)=(k, {\bf R}_k)$, we obtain $(ij|ij)$-type exchange integral as 
\begin{eqnarray}
K_{i {\bf R} j {\bf R}'}^{\sigma\rho}  = 
I_{i{\bf R},i{\bf R},j{\bf R'},j{\bf R'}}^{\sigma\rho}.  
\end{eqnarray}
With this two-center approximation, we obtain the effective Hamiltonian in Eq.~(\ref{eq:H}) characterized by \{$t_{i{\bf R} j{\bf R}'}^{\sigma\rho}$\}, \{$U_{i{\bf R} j{\bf R}'}^{\sigma\rho}$\}, \{$J_{i{\bf R} j{\bf R}'}^{\sigma\rho}$\}, and \{$K_{i{\bf R} j{\bf R}'}^{\sigma\rho}$\}. 

\section{Exchange integral in the spinor formalism}\label{app:J-and-K}
Here, we consider the relationship among the three-type exchange integrals of the Hund-type ${\cal J}^{{\rm H}}$, exchange-type ${\cal J}^{{\rm EX}}$, and pair-hopping-type ${\cal J}^{{\rm PH}}$, which are defined as 
\begin{eqnarray}
{\cal J}_{ij}^{\rm H} 
&\equiv& 
\int_V\!\!d{\bf r}\int_V\!\!d{\bf r'}
\Phi_{i\uparrow}^{\dagger}({\bf r}) \Phi_{j\uparrow}({\bf r}) 
W({\bf r,r'}) 
\Phi_{j\uparrow}^{\dagger}({\bf r'}) \Phi_{i\uparrow}({\bf r'}), \notag \\
\label{JH}
\end{eqnarray}
\begin{eqnarray}
{\cal J}_{ij}^{\rm EX} 
&\equiv& 
\int_V\!\!d{\bf r}\int_V\!\!d{\bf r'}
\Phi_{i\uparrow}^{\dagger}({\bf r}) \Phi_{j\uparrow}({\bf r}) 
W({\bf r,r'}) 
\Phi_{j\downarrow}^{\dagger}({\bf r'}) \Phi_{i\downarrow}({\bf r'}), \notag \\ 
\label{JEX} 
\end{eqnarray}
and
\begin{eqnarray} 
{\cal J}_{ij}^{\rm PH}
&\equiv&  
\int_V\!\!d{\bf r}\int_V\!\!d{\bf r'}
\Phi_{i\uparrow}^{\dagger}({\bf r}) 
\Phi_{j\uparrow}({\bf r}) 
W({\bf r,r'}) 
\Phi_{i\downarrow}^{\dagger}({\bf r'}) 
\Phi_{j\downarrow}({\bf r'}), \notag \\
\label{JPH} 
\end{eqnarray} 
respectively, where we drop the lattice-vector index ${\bf R}$ for simplicity. These integrals are related with the $J$ and $K$ matrices defined in Eqs.~(\ref{Jij}) and (\ref{Kij}) as follows:
\begin{eqnarray}
{\cal J}_{ij}^{\rm H}
&=& J_{ij}^{\uparrow\uparrow}, \\
{\cal J}_{ij}^{\rm EX}
&=& J_{ij}^{\uparrow\downarrow}, \\
{\cal J}_{ij}^{\rm PH}
&=& K_{ij}^{\uparrow\downarrow}.
\end{eqnarray}
If the Wannier function is in a scalar form, it is a real function, and then the three-type  exchange integrals are the same; ${\cal J}_{ij}^{\rm H}={\cal J}_{ij}^{\rm EX}={\cal J}_{ij}^{\rm PH}$, or equivalently, $J_{ij}=K_{ij}$, where the spin indices on the $J$ and $K$ matrix elements can be dropped, because, in the scalar case, the exchange integral is  characterized with only the spatial function. On the other hand, in the spinor case, the Wannier function is in general complex, and the three-type exchange integrals are not the same. In this appendix, we discuss the relationship among these exchange integrals. 

First, we show that the Hund-type exchange integral ${\cal J}_{ij}^{{\rm H}}$ in Eq.~(\ref{JH}) is equal to the pair-hopping-type exchange integral ${\cal J}_{ij}^{{\rm PH}}$ in Eq.~(\ref{JPH}) when there is time-reversal symmetry. By inserting the spinor component representation of Eqs.~(\ref{spinor-wannier}) and (\ref{spinor-wannier-d}) in Eq.~(\ref{JH}), the Hund-type exchange integral is written as spinor-wannier
\begin{eqnarray}
{\cal J}_{ij}^{\rm H} 
&=&
\int_V\!\!d{\bf r}\int_V\!\!d{\bf r'}
\Phi_{i\uparrow}^{\dagger}({\bf r}) \Phi_{j\uparrow}({\bf r}) 
W({\bf r,r'}) 
\Phi_{j\uparrow}^{\dagger}({\bf r'}) \Phi_{i\uparrow}({\bf r'}) \notag \\ 
&=&
\int_V\!\!d{\bf r}\int_V\!\!d{\bf r'}
\bigl(
\phi_{i\uparrow}^{u*}(\mathbf{r}) \ 
\phi_{i\uparrow}^{d*}(\mathbf{r})\bigr)
\left(\!\!\begin{array}{cc} 
    \phi_{j\uparrow}^{u}({\bf r}) \\ 
    \vspace{-0.3cm} \\
    \phi_{j\uparrow}^{d}({\bf r}) \\ 
  \end{array}\!\!\right) \notag \\ 
&\ &\ \ \ \times W({\bf r,r'}) 
\bigl(
\phi_{j\uparrow}^{u*}(\mathbf{r}') \ 
\phi_{j\uparrow}^{d*}(\mathbf{r}')
\bigr)
\left(\!\!\begin{array}{cc} 
    \phi_{i\uparrow}^{u}({\bf r}') \\ 
    \vspace{-0.3cm} \\
    \phi_{i\uparrow}^{d}({\bf r}')  \\ 
  \end{array}\!\!\right).  
  \label{H-tmp} 
\end{eqnarray}
Here, we utilize the time-reversal symmetry for the spinor as
\begin{eqnarray}
  \left(\!\!\begin{array}{cc} 
    \phi_{i\uparrow}^{u}({\bf r}) \\ 
    \vspace{-0.3cm} \\
    \phi_{i\uparrow}^{d}({\bf r}) \\ 
  \end{array}\!\!\right)
  =  
   \left(\!\!\begin{array}{cc} 
    \phi_{i\downarrow}^{d*}({\bf r}) \\ 
    \vspace{-0.3cm} \\
   -\phi_{i\downarrow}^{u*}({\bf r}) \\ 
  \end{array}\!\!\right)
 \label{TRS-up}
\end{eqnarray}
and its transpose conjugate 
\begin{eqnarray}
\bigl(
\phi_{i\uparrow}^{u*}(\mathbf{r}) \ 
\phi_{i\uparrow}^{d*}(\mathbf{r})
\bigr)
=
\bigl(
\phi_{i\downarrow}^{d}(\mathbf{r}) \ 
-\phi_{i\downarrow}^{u}(\mathbf{r})
\bigr).
  \label{TRS-upstar}
\end{eqnarray}
By inserting Eqs.~(\ref{TRS-up}) and (\ref{TRS-upstar}) in Eq.~(\ref{H-tmp}), the Hund-type exchange integral is transformed as follows:
\begin{eqnarray}
{\cal J}_{ij}^{\rm H} 
&=&
\int_V\!\!d{\bf r}\int_V\!\!d{\bf r'}
\bigl(
\phi_{i\uparrow}^{u*}(\mathbf{r}) \ 
\phi_{i\uparrow}^{d*}(\mathbf{r})\bigr)
\left(\!\!\begin{array}{cc} 
    \phi_{j\uparrow}^{u}({\bf r}) \\ 
    \vspace{-0.3cm} \\
    \phi_{j\uparrow}^{d}({\bf r}) \\ 
  \end{array}\!\!\right) \notag \\ 
&\ &\ \ \ \times W({\bf r,r'}) 
\bigl(
\phi_{j\downarrow}^{d}({\bf r}') \ 
-\phi_{j\downarrow}^{u}({\bf r}')
\bigr)
\left(\!\!\begin{array}{cc} 
    \phi_{i\downarrow}^{d*}({\bf r}') \\ 
    \vspace{-0.3cm} \\
   -\phi_{i\downarrow}^{u*}({\bf r}')  \\ 
  \end{array}\!\!\right) \notag \\ 
  &=& 
\int_V\!\!d{\bf r}\int_V\!\!d{\bf r'}
\bigl(
\phi_{i\uparrow}^{u*}({\bf r}) \ 
\phi_{i\uparrow}^{d*}({\bf r})\bigr)
\left(\!\!\begin{array}{cc} 
    \phi_{j\uparrow}^{u}({\bf r}) \\ 
    \vspace{-0.3cm} \\
    \phi_{j\uparrow}^{d}({\bf r}) \\ 
  \end{array}\!\!\right) \notag \\ 
&\ &\ \ \ \times W({\bf r,r'}) 
\bigl(
\phi_{i\downarrow}^{u*}({\bf r}') \ 
\phi_{i\downarrow}^{d*}({\bf r}')
\bigr)
\left(\!\!\begin{array}{cc} 
    \phi_{j\downarrow}^{u}({\bf r}') \\ 
    \vspace{-0.3cm} \\
    \phi_{j\downarrow}^{d}({\bf r}')  \\ 
  \end{array}\!\!\right) \notag \\    
&=&
\int_V\!\!d{\bf r}\int_V\!\!d{\bf r'}
\Phi_{i\uparrow}^{\dagger}({\bf r}) 
\Phi_{j\uparrow}({\bf r}) 
W({\bf r,r'}) 
\Phi_{i\downarrow}^{\dagger}({\bf r'}) 
\Phi_{j\downarrow}({\bf r'}). \notag \\   
\end{eqnarray}
The right hand side of the above equation is the pair-hopping-type exchange integral ${\cal J}_{ij}^{\rm PH}$ in Eq.~(\ref{JPH}), and from the view of the $J$ and $K$ matrices, we have a relationship of $J_{ij}^{\uparrow\uparrow}=K_{ij}^{\uparrow\downarrow}$.

Next, we consider the exchange-type exchange integral ${\cal J}_{ij}^{\rm EX}$. Similar to the discussion of the Hund-type exchange integral, this integral is written as
\begin{eqnarray}
{\cal J}_{ij}^{\rm EX} 
&=&
\int_V\!\!d{\bf r}\int_V\!\!d{\bf r'}
\Phi_{i\uparrow}^{\dagger}({\bf r}) \Phi_{j\uparrow}({\bf r}) 
W({\bf r,r'}) 
\Phi_{j\downarrow}^{\dagger}({\bf r'}) \Phi_{i\downarrow}({\bf r'}) \notag \\
&=&
\int_V\!\!d{\bf r}\int_V\!\!d{\bf r'}
\bigl(
\phi_{i\uparrow}^{u*}({\bf r}) \ 
\phi_{i\uparrow}^{d*}({\bf r})\bigr)
\left(\!\!\begin{array}{cc} 
    \phi_{j\uparrow}^{u}({\bf r}) \\ 
    \vspace{-0.3cm} \\
    \phi_{j\uparrow}^{d}({\bf r}) \\ 
  \end{array}\!\!\right) \notag \\ 
&\ &\ \ \ \times W({\bf r,r'}) 
\bigl(
\phi_{j\downarrow}^{u*}({\bf r}') \ 
\phi_{j\downarrow}^{d*}({\bf r}')
\bigr)
\left(\!\!\begin{array}{cc} 
    \phi_{i\downarrow}^{u}({\bf r}') \\ 
    \vspace{-0.3cm} \\
    \phi_{i\downarrow}^{d}({\bf r}')  \\ 
  \end{array}\!\!\right).  
\label{EX-tmp} 
\end{eqnarray} 
Using again the time-reversal symmetry of Eqs.~(\ref{TRS-up}) and (\ref{TRS-upstar}), ${\cal J}_{ij}^{\rm EX}$ in Eq.~(\ref{EX-tmp}) is transformed as follows:
\begin{eqnarray}
{\cal J}_{ij}^{\rm EX} 
&=&  
\int_V\!\!d{\bf r}\int_V\!\!d{\bf r'}
\bigl(
\phi_{i\uparrow}^{u*}(\mathbf{r}) \ 
\phi_{i\uparrow}^{d*}(\mathbf{r})\bigr)
\left(\!\!\begin{array}{cc} 
    \phi_{j\uparrow}^{u}({\bf r}) \\ 
    \vspace{-0.3cm} \\
    \phi_{j\uparrow}^{d}({\bf r}) \\ 
  \end{array}\!\!\right) \notag \\ 
&\ &\ \ \ \times W({\bf r,r'}) 
\bigl(
-\phi_{j\uparrow}^{d}({\bf r}') \ 
 \phi_{j\uparrow}^{u}({\bf r}')
\bigr)
\left(\!\!\begin{array}{cc} 
-\phi_{i\uparrow}^{d*}({\bf r}') \\ 
    \vspace{-0.3cm} \\
    \phi_{i\uparrow}^{u*}({\bf r}')  \\ 
  \end{array}\!\!\right) \notag \\ 
&=&
\int_V\!\!d{\bf r}\int_V\!\!d{\bf r'}
\bigl(
\phi_{i\uparrow}^{u*}({\bf r}) \ 
\phi_{i\uparrow}^{d*}({\bf r})\bigr)
\left(\!\!\begin{array}{cc} 
    \phi_{j\uparrow}^{u}({\bf r}) \\ 
    \vspace{-0.3cm} \\
    \phi_{j\uparrow}^{d}({\bf r}) \\ 
  \end{array}\!\!\right) \notag \\ 
&\ &\ \ \ \times W({\bf r,r'}) 
\bigl(
\phi_{i\uparrow}^{u*}({\bf r}') \ 
\phi_{i\uparrow}^{d*}({\bf r}')
\bigr)
\left(\!\!\begin{array}{cc} 
    \phi_{j\uparrow}^{u}({\bf r}') \\ 
    \vspace{-0.3cm} \\
    \phi_{j\uparrow}^{d}({\bf r}')  \\ 
  \end{array}\!\!\right) \notag \\ 
&=&
\int_V\!\!d{\bf r}\int_V\!\!d{\bf r'}
\Phi_{i\uparrow}^{\dagger}({\bf r}) \Phi_{j\uparrow}({\bf r}) 
W({\bf r,r'}) 
\Phi_{i\uparrow}^{\dagger}({\bf r'}) \Phi_{j\uparrow}({\bf r'}) \notag \\ 
\end{eqnarray} 
Thus, the exchange-type exchange integral ${\cal J}_{ij}^{\rm EX}$ is found to be expressed as $K_{ij}^{\uparrow\uparrow}$ in the $K$ matrix in Eq.~(\ref{Kij}), which results in $J_{ij}^{\uparrow\downarrow}=K_{ij}^{\uparrow\uparrow}$.

We summarize the structure of the $J$ and $K$ matrices as follows: 
\begin{eqnarray}
 {\bf J} 
  =\left(\begin{array}{@{\,}c|c} \\ 
  J^{\uparrow\uparrow}\!=\!{\cal J}^{{\rm H}}\!=\!{\cal J}^{{\rm PH}} & 
  J^{\uparrow\downarrow}   = {\cal J}^{{\rm EX}} \\ \\ \hline \\ 
  J^{\downarrow\uparrow}   = {\cal J}^{{\rm EX}} & 
  J^{\downarrow\downarrow}\!=\!{\cal J}^{{\rm H}}\!=\!{\cal J}^{{\rm PH}} \\ \\ 
  \end{array}\right),  
\end{eqnarray}
and 
\begin{eqnarray}
 {\bf K} 
= \left(\begin{array}{@{\,}c|c} \\ 
  K^{\uparrow\uparrow}     = {\cal J}^{{\rm EX}}  & 
  K^{\uparrow\downarrow}\!=\!{\cal J}^{{\rm PH}}\!=\!{\cal J}^{{\rm H}}\!\! \\ \\ \hline \\ 
  K^{\downarrow\uparrow}\!=\!{\cal J}^{{\rm PH}}\!=\!{\cal J}^{{\rm H}}\!\! & 
  K^{\downarrow\downarrow} = {\cal J}^{{\rm EX}}  \\ \\ 
  \end{array}\right)\!. 
\end{eqnarray}
The matrix size of ${\bf J}$ and ${\bf K}$ is $2N_w \times 2N_w$, which is composed of $N_w \times N_w$ block matrices. We note that the terms related to the $K^{\uparrow\uparrow}$ and $K^{\downarrow\downarrow}$ blocks are the terms corresponding to $a_{i\uparrow}^{\dagger} a_{j\uparrow}^{\dagger} a_{i\uparrow} a_{j\uparrow}$ and $a_{i\downarrow}^{\dagger} a_{j\downarrow}^{\dagger} a_{i\downarrow} a_{j\downarrow}$, which become zero as the result of the action on the vacuum state.
\vspace{5cm} 
\section{Single Chain Hamiltonian}\label{app:SingleChainHamiltonian}
The electronic structure of Ca$_5$Ir$_3$O$_{12}$ near the Fermi level consists of mainly $d_{xy}$/$d_{yz}$ orbitals of iridium ions, where these ions constitute the one-dimensional chains. Although the interchain couplings are important to reproduce the DFT electronic structure as shown in Fig.~\ref{band_with_without_interchain_transfer}(a) (blue curves), the single chain provides the zeroth-order approximation to capture the impact of SOI (and the broken inversion symmetry) on the electronic structure. In this appendix, we focus on the single chain of iridium and illustrate the mechanism of the gap formation below the Fermi level and spin anisotropy.

The dominant energy scales in the single-particle Hamiltonian of the single chain are the onsite spin-orbit coupling $\lambda = t_{xy\uparrow,yz\downarrow}^{\rm onsite}$, the nearest-neighbor hoppings $t_r = t_{xy\sigma,yz\sigma}^{NN}$ and $t_{\ell} = t_{yz\sigma,xy\sigma}^{NN}$. The other spin-independent (spin-dependent) matrix elements are one-order magnitude smaller in further neighbor hoppings (offsite spin-orbit couplings). 

Then, the dominant part of the single-particle Hamiltonian is given by
\begin{eqnarray}
\mathcal{H}_0 = \mathcal{H}_{0}^{\rm onsite} + \mathcal{H}_{0}^{NN},
\end{eqnarray}
where the onsite term $\mathcal{H}_{0}^{\rm onsite}$ and the nearest-neighbor term $\mathcal{H}_{0}^{\rm NN}$
are defined as
\begin{widetext}
\begin{eqnarray}
\mathcal{H}_0^{\rm onsite} =
\sum_{{\bf R}}
\left(
\begin{array}{cccc}
a_{xy\uparrow{\bf R}}^{\dagger} &
a_{yz\uparrow{\bf R}}^{\dagger} &
a_{xy\downarrow{\bf R}}^{\dagger} &
a_{yz\downarrow{\bf R}}^{\dagger} \\
\end{array}
\right)
\left(
\begin{array}{cccc}
\varepsilon - \mu  & -d & 0 & \lambda \\
-d &  \varepsilon - \mu  & -\lambda & 0  \\
0 & -\lambda & \varepsilon - \mu & -d \\
\lambda & 0 & -d & \varepsilon - \mu \\
\end{array}
\right)
\left(
\begin{array}{c}
a_{xy\uparrow{\bf R}} \\
a_{yz\uparrow{\bf R}} \\
a_{xy\downarrow{\bf R}} \\
a_{yz\downarrow{\bf R}} \\
\end{array}
\right),
\end{eqnarray}
and
\begin{eqnarray}
\mathcal{H}_0^{NN} &=& 
\sum_{{\bf R}}
\left(
\begin{array}{cccc}
a_{xy\uparrow{\bf R}}^{\dagger} &
a_{yz\uparrow{\bf R}}^{\dagger} &
a_{xy\downarrow{\bf R}}^{\dagger} &
a_{yz\downarrow{\bf R}}^{\dagger} \\
\end{array}
\right)
\left(
\begin{array}{cccc}
t_0 & t_r & 0 & 0 \\
t_{\ell} & t_0 & 0 & 0  \\
0 & 0 & t_0  & t_r \\
0 & 0 & t_{\ell} & t_0 \\
\end{array}
\right)
\left(
\begin{array}{c}
a_{xy\uparrow{\bf R}+{\bf R}_c} \\
a_{yz\uparrow{\bf R}+{\bf R}_c} \\
a_{xy\downarrow{\bf R}+{\bf R}_c} \\
a_{yz\downarrow{\bf R}+{\bf R}_c} \\
\end{array}
\right)\nonumber\\
&&+
\sum_{\bf R}
\left(
\begin{array}{cccc}
a_{xy\uparrow{\bf R}}^{\dagger} &
a_{yz\uparrow{\bf R}}^{\dagger} &
a_{xy\downarrow{\bf R}}^{\dagger} &
a_{yz\downarrow{\bf R}}^{\dagger} \\
\end{array}
\right)
\left(
\begin{array}{cccc}
t_0  & t_{\ell} & 0 & 0 \\
t_r & t_0  & 0 & 0  \\
0 & 0 & t_0  & t_{\ell} \\
0 & 0 & t_r & t_0 \\
\end{array}
\right)
\left(
\begin{array}{c}
a_{xy\uparrow{\bf R}-{\bf R}_c} \\
a_{yz\uparrow{\bf R}-{\bf R}_c} \\
a_{xy\downarrow{\bf R}-{\bf R}_c} \\
a_{yz\downarrow{\bf R}-{\bf R}_c} \\
\end{array}
\right),
\end{eqnarray}
where, $d=t_{xy\sigma,yz\sigma}^{\rm onsite}$, $t_0 = t_{xy\uparrow,xy\uparrow}^{NN}=t_{xy\downarrow,xy\downarrow}^{NN} \simeq t_{yz\uparrow,yz\uparrow}^{NN}=t_{yz\downarrow,yz\downarrow}^{NN}$, $\varepsilon$ denotes the onsite energy for the $d_{xy}$ and $d_{yz}$ orbitals, $\mu$ is the chemical potential, and ${\bf R}_c$ is the lattice vector along the $c$ axis. Here, we neglect the matrix elements smaller than 10 meV in $\mathcal{H}_0^{\rm onsite}$ and $\mathcal{H}_0^{NN}$. 

To make the nature of $\mathcal{H}_0$ transparent, we perform the Fourier transformation of $\mathcal{H}_0$ by introducing the Pauli matrices for the spin and orbital degrees of freedom, $\sigma^{\alpha}$ and $\tau^{\alpha}$
($\alpha=0,x,y,z$), respectively, as
\begin{eqnarray}
\mathcal{H}_0
&=&
\sum_{\bf k}
\left\{
\left[
\varepsilon - \mu + 2 t_0 \cos \left({\bf k} \cdot {\bf R}_c \right)
\right]\sigma^0 \otimes \tau^0
+
\left[ -d + (t_r +t_{\ell})\cos \left({\bf k} \cdot {\bf R}_c \right) \right]
\sigma^0 \otimes \tau^x
\right.
\nonumber\\
&&-
\left.
(t_r -t_{\ell})\sin \left({\bf k} \cdot {\bf R}_c \right) \sigma^0 \otimes \tau^y
-\lambda \sigma^y \otimes \tau^y\right\},
\end{eqnarray}
where $\otimes$ denotes the Kronecker product of two matrices.
Then, the band dispersions for the conduction bands $E_{\rm c}^{\pm}$ and the valence bands $E_{\rm v}^{\pm}$
are obtained as,
\begin{eqnarray}
E_{\rm c}^{\pm} &=& 
\varepsilon - \mu + 2 t_0 \cos \left({\bf k} \cdot {\bf R}_c \right)
+\sqrt{\left[ -d + (t_r + t_{\ell})\cos \left({\bf k} \cdot {\bf R}_c \right) \right]^2
+
\left[\pm\lambda + (t_r - t_{\ell})\sin \left({\bf k} \cdot {\bf R}_c \right) \right]^2},\\
E_{\rm v}^{\pm} &=&
\varepsilon - \mu + 2 t_0 \cos \left({\bf k} \cdot {\bf R}_c \right)
 -\sqrt{\left[ -d + (t_r + t_{\ell})\cos \left({\bf k} \cdot {\bf R}_c \right) \right]^2
+
\left[\pm\lambda + (t_r - t_{\ell})\sin \left({\bf k} \cdot {\bf R}_c \right) \right]^2}.
\end{eqnarray}
\end{widetext}

The gap among conduction and valence bands is simply estimated as follows. Because the ratio $d/(t_r + t_{\ell})$ ($\sim 0.07$ for SO-GGA) is small and negligible, $|E_{\rm c}^{\pm} - E_{\rm v}^{\pm}|$ and $|E_{\rm c}^{\pm} - E_{\rm v}^{\mp}|$ show minima at ${\bf k} \cdot {\bf R}_c \sim \pm \pi/2$, where $ -d + (t_r+t_{\ell})\cos \left({\bf k} \cdot {\bf R}_c \right) \simeq 0$. Then the gap is estimated by $2|\pm \lambda + (t_r -t_{\ell})|$ and, thus, is determined by the combination of the spin-orbit coupling $\lambda$ and broken inversion symmetry quantified by $|t_r -t_{\ell}|$. 

The single-particle states around the conduction band minima at ${\bf k} \cdot {\bf R}_c \sim \pm \pi/2$ show
easy-axis anisotropy characterized by the $y$ component of the spin-orbit coupling, as follows. Around the conduction band minuma,
the single-particle Hamiltonian is dominated by,
\begin{eqnarray}
\mp (t_r -t_{\ell}) \sigma^0 \otimes \tau^y -\lambda \sigma^y \otimes \tau^y.
\end{eqnarray}
The above term is diagonalized by the eigenstates of $\sigma^y \otimes \tau^y$.

The $y$ component of the Pauli matrices for the orbital degrees of freedom is proportional to a projection of the $y$ component of the effective angular momentum, which is denoted by $\ell_y$, in the $t_{2g}$ manifold onto the two-dimensional $d_{xy}/d_{yz}$ manifold: $\lambda \sigma^y \otimes \tau^y$ is equivalent to $\lambda \sigma^y \otimes \ell^y$.
In the $t_{2g}$ manifold expanded by the maximally localized Wannier orbitals, the spin-orbit coupling is given by $-\lambda_{t_{2g}}\vec{\ell}\cdot\vec{\sigma}+\delta \lambda_{t_{2g}} \ell_y \sigma_y + \lambda_{t_{2g}}'(\ell_z \sigma_x + \ell_x \sigma_z)$, where $\lambda_{t_{2g}}(>0)$ is the spin-orbit coupling in the $t_{2g}$ manifold, and $\delta \lambda_{t_{2g}}$ and $\lambda_{t_{2g}}'$ are anisotropic couplings due to the crystal fields.
When the $zx$ orbital is fully occupied, only the $y$ component of the above spin-orbit coupling remains relevant. We note that, by reversing the sign of the effective angular momentum $\ell_y$, we obtain the $y$ component of the angular momentum $L_y (= - \ell_y)$ in the spherical environment.
Thus, $-\lambda \sigma_y \ell_y$ is nothing but the $y$ component of the $LS$ coupling. 

The eigenstates of $\sigma^y \otimes \tau^y$ are labeled by the $y$ component of the effective total angular momentum, $J^y_{\rm eff}=\pm1/2, \pm 3/2$.
In the broken time-reversal symmetry phases, the conduction electrons may induce the total angular momentum along the local $y$ axis due to the the $y$ component of the $LS$ coupling. 

%
\end{document}